\documentclass[11pt]{article}
\usepackage{amsmath,amssymb,color,epsfig,cite}
\usepackage{setspace}
\usepackage{comment}

\usepackage{amsfonts}
\usepackage{slashed}
\usepackage{physics}

\newcommand{\hoch}[1]{$\, ^{#1}$}

\makeatletter
\@addtoreset{equation}{section}
\makeatother

\newcommand{\be}{\begin{equation}}
\newcommand{\ee}{\end{equation}}
\newcommand{\bea}{\setlength\arraycolsep{2pt} \begin{eqnarray}}
\newcommand{\eea}{\end{eqnarray}}

\def\dd{{\slashed \delta}}

\newcommand{\bpm}{\begin{pmatrix}}
\newcommand{\epm}{\end{pmatrix}}

\def\0{{\sst{(0)}}}
\def\1{{\sst{(1)}}}
\def\2{{\sst{(2)}}}
\def\3{{\sst{(3)}}}
\def\4{{\sst{(4)}}}
\def\5{{\sst{(5)}}}
\def\6{{\sst{(6)}}}
\def\7{{\sst{(7)}}}
\def\8{{\sst{(8)}}}

\def\sst#1{{\scriptscriptstyle #1}}

\usepackage{mathrsfs}

\usepackage{enumitem} 
\usepackage{xcolor}
\usepackage{hyperref}
\hypersetup{hidelinks}

\begin{document}

\begin{flushright}
\end{flushright}

\vspace{15pt}

\begin{center}
{\Large {\bf Dual Charges for AdS Spacetimes and the First Law of Black Hole Mechanics}}

\vspace{15pt}

\vspace{15pt}
{\bf Mahdi Godazgar and Simon Guisset}

\vspace{10pt}

\hoch{1} {\it School of Mathematical Sciences,
Queen Mary University of London, \\
Mile End Road, E1 4NS, UK.}

 \vspace{15pt}
 
\today

\vspace{20pt}

\underline{ABSTRACT}
\end{center}

\noindent We apply the recent derivations of dual charges in asymptotically flat spacetimes to asymptotically locally AdS spacetimes.  In contrast to the results in the flat case, in the AdS case with a Dirichlet boundary the dual charge contribution vanishes at the leading order.  However, by focusing on the Taub-NUT-AdS solution, we show that nevertheless, more generally, the dual charge is non-vanishing and corresponds to the NUT parameter.  We propose a complex first law of black mechanics in the presence of NUT charges that is inspired by the naturally complex nature of the charges derived using Hamiltonian methods.

\thispagestyle{empty}

\vfill
\noindent Emails:\hspace{-1mm} m.godazgar@qmul.ac.uk, s.guisset@qmul.ac.uk

\pagebreak

\tableofcontents

\section{Introduction}

The study of charges in asymptotically (locally) anti-de Sitter (AdS) backgrounds has a rich history \cite{Abbott:1981ff, BH86} that has taken on a particular importance in the context of the AdS/CFT correspondence \cite{Maldacena}. This correspondence, formulated in a stringy framework, establishes a physical duality between $(D+1)$-dimensional theories of quantum gravity with negative cosmological constant and a $D$-dimensional conformal field theory at the boundary.

In this paper, we revisit the study of AdS charges in light of recent developments in the flat case \cite{Godazgar:2018qpq, Godazgar:2018dvh, Godazgar:2020gqd, Godazgar:2020kqd}.  In Ref.\ \cite{Godazgar:2020gqd} it is argued that in order to have access to all possible gravitational charges, one must include in the gravitational lagrangian any terms that do not contribute to the equations of motion.  One simple such term is the Holst term \cite{Holst:1995pc}, whose equations of motion give the algebraic Bianchi identity and are therefore trivial.  Nevertheless, the inclusion of such a term leads to non-trivial charges, which correspond \cite{Godazgar:2020kqd} to the dual charges of Refs.\ \cite{Godazgar:2018qpq, Godazgar:2018dvh}.  The main question we would like to address here is what happens if we apply this idea to asymptotically (locally) AdS backgrounds?

There are many methods to derive charges/Hamiltonians in AdS backgrounds.  These include the covariant phase space formalism \cite{phaseconserved1, phaseconserved2, Aros:1999id, Hollands:2005wt, Marolf2014}, the asymptotic study of Brown-Henneaux \cite{BH86, Papadimitriou:2005ii, Odak:2021axr}, conformal methods (AMD) \cite{Ashtekar_1984, Ashtekar2000} and cohomological methods \cite{Abbott:1981ff, Barnich:2001jy, Barnich:2003qn, Barnich:2004uw}.  The study of charges has been motivated amongst other things by an understanding of the first law of black hole mechanics \cite{Frodden:2017qwh, Durka:2011yv, Durka:2011zf, Johnson:2014xza, Hennigar2019, Peng:2020cfy}.

Here, as in Ref.\ \cite{Godazgar:2020gqd}, we will use the covariant phase space formalism,\footnote{While, more clear in the covariant phase space formalism, dual charges can also be derived using cohomological methods \cite{Oliveri:2020xls}.} in which charges are derived as hamiltonians associated with asymptotic symmetry transformations on the covariant phase space. These transformations are dictated by the boundary conditions one prescribes at the AdS boundary. One typically introduces Dirichlet boundary conditions (fixing the induced metric at the boundary) or Neumann boundary conditions (fixing the energy momentum tensor $T^{jk}$) \cite{CompereRuzziconi, Fiorucci:2020xto}. In four dimensions, the first conditions give the group of isometries of AdS$_4$ while the second conditions gives an empty asymptotic group. One can also introduce ``mixed" boundary conditions as in \cite{CompereRuzziconi} giving a group of the form $\mathbb{R}\oplus \mathcal{A}$ with $\mathcal{A}$ a group of area-preserving transformations. 

We will in this work find such charges for the Einstein-Palatini-Holst action. This action will give two sets of distinct charges: the usual charges, that will be referred to as the ``electric charges" while the second set will be referred to as the ``dual charges". The latter come from the Holst term in the action. Dual asymptotic charges may be thought of as the generalisation of NUT charges in the same way that the usual asymptotic charges are the generalisation of mass.  In turn NUT charges may be viewed as the gravitational analogues of magnetic charges.  In fact this relation can be made quite explicit in a characteristic value formulation of the Einstein equation \cite{MahdiTaubNUT}.

In the flat context, the existence of dual charges is important for a Hamiltonian interpretation of Newman-Penrose charges \cite{Godazgar:2018vmm, Godazgar:2018dvh} and the consistency of the action of BMS charges on phase space \cite{Godazgar2019}.  While dual charges in the spirit of Ref.\ \cite{Godazgar:2020gqd} have not been studied in the AdS context prior to this work, there is a rich literature on NUT or more generally magnetic charges in the AdS backgrounds \cite{Hawking:1998ct, Miskovic:2009bm, Johnson:2014xza,Araneda:2016iiy, Aros:2017wun,Giribet:2018hck,Giribet:2020aks, Arratia:2020hoy, Ciambelli:2020qny, Flores-Alfonso:2020nnd, Mann:2020wad, Andrianopoli:2021qli, Rodriguez:2021hks}.  For the most part, the interest in magnetic masses is with relation to a consistent formulation of black holes thermodynamics.  In particular, \cite{Hawking:1998ct} argues that Misner strings do contribute to the entropy.  In \cite{Araneda:2016iiy}, it is argued that the Taub-NUT solution may be interpreted most appropriately as a ground state in the regularised theory.  This is because the presence of a NUT charges causes the total Noether charge to vanish for this solution. From a holographic perspective, the existence of magnetic charges is useful for studying the richness of the boundary theory.  Moreover, such solutions always lift to solutions of M-theory and are as such consistent \cite{Martelli:2012sz}.

We will find that the dual charges defined following the prescription of \cite{Godazgar:2020gqd} vanish on the boundary for asymptotically AdS spacetimes with Dirichlet boundary conditions.  To be precise, we find that the dual charges are of the form: 
\begin{equation}
    \dd_\xi \tilde{H} = \mathcal{O}(z),
\end{equation}
where $z$, defined in section \ref{definition and action}, parametrises the distance from the boundary.  More generally, however, dual charges are possible.  In particular, they are non-vanishing for backgrounds with global dual or NUT charges, such as the Taub-NUT-AdS solution, which is parametrised by a mass parameter $m$ and NUT parameter $n$.  Deriving the charges for this solution, as expected, one gets the usual energy $E=m$ in the electric part, plus an angular momentum contribution from the Misner string that can be taken to be zero if the wire singularity is equally distributed between the north and south poles of the sphere. The dual charge is given by 
\begin{align}
    \tilde{E} = n\left(1-\frac{4\Lambda}{3}n^2\right),
\end{align}
where $\Lambda$ is the cosmological constant.  The existence of dual global charges naturally leads to an investigation of the first law of black holes in the presence of NUT charges within the context of the covariant phase space formalism; a subject of recent interest \cite{Johnson:2014xza, Hennigar2019, Durka:2019ajz, Ciambelli:2020qny,  Rodriguez:2021hks, Frodden:2021ces}.  In contrast to previous literature, we take the interpretation of the NUT charge as a dual mass \cite{Ramaswamy} seriously.  This leads us to view the complex combination of the mass and NUT charge (see also \cite{Freidel:2021qpz, Freidel:2021ytz})
\begin{equation}
\mathcal{M} = m - i n
\end{equation}
as a generalised mass that satisfies its own first law.

In section \ref{sec:tetrad}, we will review the first-order tetrad formalism. The tetrad formalism allows one to write more general terms in the action, which do not contribute to the equations of motion.  In section \ref{FGexpansion}, we introduce \emph{Asymptotically AdS} (aAdS) and \emph{Asymptotically Locally AdS} (alAdS) backgrounds and explain the difference between them.  Moreover, we define the Fefferman-Graham expansion of the metric of alAdS spacetimes and review the symmetries of the backgrounds.  In section \ref{sec:charges}, we apply the covariant phase space formalism to AdS backgrounds and derive the Hamiltonians in this context.  An important element to consider is the role of the internal Lorentz transformations, which add further gauge degrees of freedom. For the charges to be Lorentz invariant, one has to introduce compensating Lorentz transformations using the prescription outlined in Ref.\ \cite{Godazgar:2022foc}.  In section \ref{sec:alAdScharges} we compute the charges for alAdS spacetimes, verifying that the standard charges correspond to the well-known Brown-York charges and finding that there are no dual charges. In the final section \ref{sec:NUT}, we focus on the Taub-NUT-AdS solution. We find that in this case the dual charge is non-zero.  We propose a first law of black hole thermodynamics. 

\paragraph{Notation:}
Lowercase Latin letters denote general tangent indices ($a,b,\dots =0,\dots 3$) and lowercase hatted Latin letters denote the tangent subspace ($\hat{a},\hat{b},\dots = 0,\dots,2$). Similarly, Greek letters $\mu,\nu,\dots$ denote general spacetime indices and Latin letters $i,j,\ldots$ denote transverse directions.

\section{First-order tetrad formalism} \label{sec:tetrad}

General Relativity is usually written in terms of spacetimes fields (such as the metric $g_{\mu\nu}$, the stress-energy tensor $T_{\mu\nu}$, ...) with dynamics dictated by the Einstein-Hilbert action: 
\begin{equation}
S_{EH}\left[g\right] = \kappa \int_M \sqrt{\vert g\vert} \left( R\left[g\right] -2\Lambda\right) + S_M,
\end{equation}
where $\kappa = \frac{1}{16\pi G}$, $\Lambda$ is the cosmological constant and $S_M$ the action for matter. This formalism, although intuitive, turns out to be incompatible with the addition of fermions interacting with the metric. The reason is actually rather easy to understand: for the same reason that a spinor can be seen as the ``square root" of a spacetime vector, we will define the frame field\footnote{Also called a vielbein in $n$ dimensions or vierbein and tetrad in $4d$.} $e^a$ as the ``square root of metric": 
\begin{equation}
\eta_{ab}e^a_\mu e^b_\nu = g_{\mu\nu}.
\end{equation}
These fields can formally be seen as Lorentz-valued spacetime one-forms and are thus fixed up to a Lorentz transformation in $SO(1,n-1)$. Such objects are transported in a non-trivial way under parallel transport and one needs to define a covariant derivative ``seeing" the Lorentz indices: \begin{equation}
\mathcal{D} e^a = d e^a + \omega^{a}_{\,\,b}\wedge  e^b,
\end{equation}
where $\omega^{ab}$ is the connection on the tangent space obeying the following constraint: 
\begin{equation}
\mathcal{D}_\mu \eta^{ab}=0,
\end{equation}
which, in particular, forces $\omega^{ab}$ to be antisymmetric. We also define the Riemann curvature two-form as the covariant derivative of the spin connection: 
\begin{equation}\label{curvature}
\mathcal{R}^{ab}=d\omega^{ab}+ \omega^{a}_{\, \,c} \wedge \omega^{cb}.
\end{equation}
Using the expression of the 2-form curvature \eqref{curvature} above, one can at this stage easily prove the following differential Bianchi identity that will be heavily used in the following: 
\begin{equation}\label{Bianchi}
    d\mathcal{R}^{ab} = \mathcal{R}^a{}_c\wedge \omega^{cb} - \omega^a{}_c \wedge \mathcal{R}^{cb}.
\end{equation}
This identity can also be understood as the vanishing of the covariant derivative of the curvature: 
\begin{equation}
D\mathcal{R} = d\mathcal{R} - \left[ \mathcal{R},\omega\right]=0,
\end{equation}
where $\left[ \mathcal{R},\omega \right]=\mathcal{R}\wedge \omega - \omega \wedge \mathcal{R}$.

In this formalism, General Relativity is described as a gauge theory where $\{\omega,\mathcal{R}\}$ are the $\mathfrak{sl}(2,\mathbb{C})$ gauge connections and curvatures, respectively. The frame field $e$ is thus a section on the associated vector bundle.  It is easy to see that the Einstein-Hilbert action can be written in terms of these new fields as: 
\begin{equation}
    S_P\left[e,\omega\right] = \frac{\kappa}{2}\varepsilon_{abcd}\int_M \mathcal{R}^{ab}\wedge e^c\wedge e^d + \frac{\ell^{-2}}{2} e^a\wedge e^b \wedge e^c \wedge e^d,
\end{equation}
where we have introduced the AdS length $\ell^{2}=-\frac{3}{\Lambda}>0$.  The first order formalism, in which we shall work, is defined by taking $\{e,\omega\}$ as independent fields in contrast to the second order formalism in which $\{e\}$ is the only dynamical field.  This means that there is an equation of motion associated with both these fields:
\begin{align}\label{equations of motion}
    &\varepsilon_{abcd}\left(\mathcal{R}^{ab} + \ell^{-2} e^a\wedge e^b\right) \wedge e^c=0,\\
    &de^a + \omega^a_{\, \; b}\wedge e^b=0.
\end{align}
The first equation, which corresponds to the equation of motion of the vierbein is equivalent to the usual Einstein equation in metric form, while the second one, corresponding to the equation of motion of the spin connection, implies the vanishing of the torsion tensor $T^a = De^a$.  Note that in contrast to the second order formalism where we take $\omega$ as being fixed by $e$ from the algebraic equation $T^a=0$, in the first order formalism, the vanishing of the torsion is a consequence of the equations of motion.

The first term in the action can also be written as: 
\begin{equation}
S_P\left[e,\omega\right] = \kappa \int_M \mathcal{R}\wedge \star\left( e\wedge e\right),
\end{equation}
which provides a guide for  how we could add another term to this action by drawing a parallel with electromagnetism. Starting with the usual Maxwell action,
\begin{equation}
S_M = e\int_M F\wedge \star F,
\end{equation}
it is well-known that one can add to this action a topological term
\begin{equation}
S_\theta = \frac{\theta}{2\pi} \int_M F\wedge F
\end{equation}
whose equations of motion is the Bianchi identity, which is trivially satisfied.  Such a term in electromagnetism has important consequences in the quantum theory and when non-zero, violates CP-symmetry.  It can be analogously constructed in gravity and is called the Holst term \cite{Holst:1995pc}:\footnote{More generally, one considers the Nieh-Yan topological invariant (see \cite{Godazgar:2020gqd} for more details):
$$
S_{NY} =\kappa\, i \,  \lambda \int_M \left( \mathcal{R}^{ab}\wedge e_a \wedge e_b - T^a \wedge T_a \right).
$$}
\begin{equation}
S_H =\kappa\, i \,  \lambda \int_M \mathcal{R}^{ab}\wedge e_a \wedge e_b,
\end{equation}
where constant $\lambda$ is usually called the Immirzi parameter.  There is \textit{a priori} no reason not to add this term. The equations of motion from this term, assuming $T^a=0$, correspond to the algebraic Bianchi identity, which is trivially zero on-shell. However such a term can give non-trivial asymptotic charges \cite{Godazgar:2020gqd, Godazgar:2020kqd} by modifying the symplectic structure of the theory such as in asymptotically flat spacetimes where it leads to dual BMS charges \cite{Godazgar:2018dvh,Godazgar:2018qpq}. The study of these dual charges in the AdS context will be the focus of this work and we will therefore consider the full Palatini-Holst action: 
\begin{equation}\label{PH Action}
    S_{PH} = \kappa P_{abcd} \int_M \left(\mathcal{R}^{ab}\wedge e^c \wedge e^d +\frac{\ell^{-2}}{2}e^a\wedge e^b\wedge e^c\wedge e^d\right) + \Omega\left[ e\vert_{\partial M},\omega\vert_{\partial M}\right],
\end{equation}
where 
\begin{equation} \label{P}
P_{abcd}=\frac{1}{2}\varepsilon_{abcd} + i\lambda \eta_{a\left[c\right.}\eta_{\left. d\right]b}
\end{equation}
and $\Omega$ is a boundary term that is a functional of the fields at infinity.\footnote{We will see in section \ref{definition and action} the significance of this boundary term.} Before proceeding, we make two remarks: 
\begin{enumerate}
    \item The tensor $P_{abcd}$ obeys the following symmetry relations: $P_{abcd}=P_{cdab}$, $P_{abcd}=-P_{bacd}$. It thus can be viewed as a $6\times 6$ symmetric tensor $P_{[ab][cd]}$.  Such a tensor is invertible for $\lambda \neq \pm 1$: 
    \begin{equation}
    P_{abcd}^{-1}=\frac{1}{2(\lambda^2-1)}\left( \varepsilon_{abcd} - 2i\lambda \eta_{a\left[c\right.}\eta_{\left.d\right]b}\right).
    \end{equation}
    \item The boundary term plays an important role in the covariant phase space formalism.  It will be chosen such that the spacetime is indeed asymptotically AdS. 
\end{enumerate}

\section{alAdS spacetimes and the Fefferman-Graham expansion}\label{FGexpansion}

\subsection{Definition, action and topological renormalization}\label{definition and action}

We recall here that AdS$_4$ is the unique maximally symmetric solution to the vacuum Einstein equations with a negative cosmological constant $\Lambda = -3/\ell^2$. It is known that the Riemann curvature of a maximally symmetric spacetime can be written explicitly as
\begin{equation}
R_{\mu\nu\gamma\lambda} + \ell^{-2} \left( g_{\mu\gamma}g_{\nu\lambda} - g_{\mu\lambda}g_{\nu\gamma}\right)=0.
\end{equation}
Note that this condition can be written in terms of the Riemann curvature 2-form $\mathcal{R}^{ab}$ as
\begin{equation} \label{ads}
\bar{\mathcal{R}}^{ab}\equiv\mathcal{R}^{ab} + \ell^{-2} e^a\wedge e^b=0.
\end{equation}
The curvature 2-form $\bar{\mathcal{R}}^{ab}$ also satisfies the differential Bianchi equation:
\begin{equation}\label{Bianchib}
    d\bar{\mathcal{R}}^{ab} = \bar{\mathcal{R}}^a{}_c\wedge \omega^{cb} - \omega^a{}_c \wedge\bar{\mathcal{R}}^{cb}.
\end{equation}
Inspired by equation \eqref{ads}, we define alAdS spacetimes \cite{Ashtekar_1984, Ashtekar2000} (see also \cite{Marolf2014}) to be a solution of (\ref{equations of motion}) such that its metric obeys the fall-off condition 
\begin{equation} \label{alAdScon}
\bar{\mathcal{R}}^{ab}\vert_{\mathcal{I}}=0
\end{equation}
with $\mathcal{I}$ the conformal boundary of $\mathcal{M}$.\footnote{In contrast, we define an asymptotically AdS spacetime to be alAdS spacetime together with a Dirichlet boundary condition that fixes the induced boundary $\mathbb{R}\times S^2$.} It was proven by Fefferman-Graham \cite{AST_1985__S131__95_0} that one can always find a set of coordinates $(z,x^a)$ at least near the conformal boundary $\mathcal{I}$ such that $\lbrace z=0\rbrace\subset\mathcal{I}$ and the metric is of the form
\begin{equation}\label{FGexpansion1}
            g = \frac{1}{z^2}\Big( \ell^2dz^2 + \gamma_{ij}(z,x^a)dx^idx^j\Big),
\end{equation}
where we have an expansion for metric $\gamma$ of the form
\begin{equation}
     \gamma = \gamma^{(0)} + z^2 \gamma^{(2)} + z^3 \gamma^{(3)} + \mathcal{O}(z^4).
\end{equation}
We then locally define the boundary as the Lorentzian manifold $(\mathcal{I},\gamma^{(0)})$.  It can be shown that the equations of motion fix $\gamma^{(2)}$ and the trace of $\gamma^{(3)}$:
\begin{equation}
\gamma^{(2)}_{ij} = -R^{(0)}_{ij} + \frac{1}{4} R^{(0)} \gamma^{(0)}_{ij}, \qquad  \gamma^{(0)\, ij} \, \gamma^{(3)}_{ij} = 0,
\end{equation}
where $R^{(0)}_{ij}$ and $R^{(0)}$ are the Ricci tensor and scalar of $\gamma^{(0)}$, respectively.

Such spacetimes are in fact very general, and the question of whether they give rise to well-posed problems is subtle. A sufficient condition for having a well-posed alAdS problem is to require the action to have an extremum when one imposes $\bar{\mathcal{R}}^{ab}=0$ at $\mathcal{I}$. 

Following \cite{Aros:1999id}, we consider a variation of the action (\ref{PH Action}), which gives
\begin{equation}
\delta S = 2\kappa\int_M P_{abcd}\left[ \bar{\mathcal{R}}^{ab}\wedge e^c \wedge \delta e^d - T^a\wedge e^b \wedge\delta \omega^{cd}\right] + \int_{\mathcal{I}} \theta(e,\omega,\delta e, \delta \omega),
\end{equation}
where $\theta(e,\omega,\delta e, \delta \omega)=\kappa P_{abcd}e^a\wedge e^b\wedge \delta \omega^{cd}+\delta \Omega$ is called the presymplectic potential. Using the equations of motion and the algebraic Bianchi identity, the first integral vanishes.  Therefore, in order to have an extremum for any alAdS spacetime, one must choose $\Omega$ so that $\theta$ is proportional to $\bar{\mathcal{R}}^{ab}$ and thus vanishes on-shell.  It is clear that this is achieved by choosing $\Omega$ such that
\begin{equation}
\delta\Omega = \kappa \ell^2 P_{abcd}\int_{\mathcal{I}} R^{ab}\wedge \delta \omega^{cd}.
\end{equation}
In this case,
\begin{equation}
\theta(e,\omega,\delta e,\delta \omega)=\kappa \ell^2 P_{abcd} \overline{\mathcal{R}}^{ab}\wedge \delta \omega^{cd}\vert_\mathcal{I},
\end{equation}
which does indeed vanish using condition \eqref{alAdScon}. Observe that $\delta\Omega$ is nothing but the variation of a quantity that may be viewed as a generalised Euler density,
\begin{equation}
    P_{abcd}\int_\mathcal{I} \mathcal{R}^{ab}\wedge\delta \omega^{cd}=\frac{1}{2}\delta \left( P_{abcd} \int_M \mathcal{R}^{ab}\wedge\mathcal{R}^{cd}\right).
\end{equation}
Substituting this expression for the boundary term $\Omega$ in action \eqref{PH Action} gives the full alAdS action, up to a constant,
\begin{equation} \label{action}
    S\left[ e,\omega\right]=\frac{\kappa \ell^2}{2} \int_M P_{abcd}\bar{\mathcal{R}}^{ab}\wedge \bar{\mathcal{R}}^{cd}.
\end{equation}
In summary, in order to define a well-posed problem, we have added simply a higher-order topological term proportional to $\ell^2$ to the original action. Furthermore, as we shall find below, this addition actually renormalizes the Euclidean action and makes the boundary charges well-defined \cite{Aros:1999id, Ciambelli:2020qny}.  Such a procedure is known as a \emph{topological renormalization}.

\subsection{Symmetries of alAdS spacetimes} \label{sec:symm}

We work with Fefferman-Graham coordinates $(z,x^i)$ valid in some neighbourhood of the boundary $\left\{ z=0\right\}$, where the metric takes the form (\ref{FGexpansion1}). 

From the form of the metric, we choose a canonical frame $(e^{\hat{a}},e^3)$ of the form 
\begin{equation}\label{expansionvierbeins}
    e^{\hat{a}}_k dx^k = \frac{1}{z}\left(e^{(-1)\hat{a}} + z^2 e^{(1)\hat{a}} + z^3 e^{(2)\hat{a}} \right) +\mathcal{O}(z^4), \qquad
    e^3_z dz = \frac{\ell}{z} \,dz,
\end{equation}
where
\begin{equation}
\gamma^{(0)}_{ij} = e^{(-1)\hat{a}}_{(i} e^{(-1)\hat{b}}_{j)} \eta_{\hat{a} \hat{b}}, \quad 
\gamma^{(2)}_{ij} = 2 e^{(-1)\hat{a}}_{(i} e^{(1)\hat{b}}_{j)} \eta_{\hat{a} \hat{b}}, \quad 
\gamma^{(3)}_{ij} = e^{(-1)\hat{a}}_{(i} e^{(2)\hat{b}}_{j)} \eta_{\hat{a} \hat{b}}.
\end{equation}

The inverse vierbeins are given by
\begin{align}
   & e_{\hat{a}}^k \partial_k =z\left\{e^{(-1)k}_{\hat{a}}  +z^2\left(e^{(1)k}_{\hat{a}}- \gamma^{(2)ik}e^{(-1)}_{\hat{a} i} \right) + z^3\left(e^{(2)k}_{\hat{a}}- \gamma^{(3)ik}e^{(-1)}_{\hat{a} i}\right)+\mathcal{O}(z^4) \right\} \partial_k, \notag \\[2mm]
    &e_3^z\partial_z = \frac{z}{\ell}\partial_z,
\end{align}
where we have used the convention that all $\hat{a},\hat{b},\ldots$ indices are lowered/raised with $\eta_{\hat{a}\hat{b}}$ and its inverse and all $i,j,\ldots$ indices are raised and lowered with $\gamma^{(0)}_{ij}$ and its inverse.  Thus, for example,
\begin{equation}
e^{(-1)i}_{\hat{a}} = \gamma^{(0)ij}e_j^{(-1)\hat{b}} \eta_{\hat{a}\hat{b}}.
\end{equation}

We will be interested in isometries of alAdS backgrounds, namely, diffeomorphisms that keep the form of the Fefferman-Graham expansion unchanged. In particular, we require that the action of a diffeomorphism $\xi$ be such that
\begin{equation}
    \delta_{\xi} g_{zz}=0=\delta_{\xi} g_{zi}.
\end{equation}
Or in terms of coordinates: 
\begin{align}
    \partial_z \left( \frac{\xi^z}{z}\right)=0, \qquad 
    \partial_z \xi ^k \gamma_{ki}+\ell^2\partial_i \xi^z=0.
\end{align}
These equations may be integrated easily to give
\begin{align}
    &\xi^z(z, x^i) = z\hat{\xi}^z(x^i), \notag \\ \label{expansion}
    &\xi^i(z,x^i) = \hat{\xi}^i(x^i) -\ell^2 \partial_j \hat{\xi}^z \int_0^z s\gamma^{ij}(s,x^i)ds=\hat{\xi}^i(x^i)-\frac{\ell^2}{2}z^2\gamma^{(0)ij}\partial_{j}\hat{\xi}^z(x^i)+\mathcal{O}(z^4),
\end{align}
where $\hat{\xi}^\mu$ is simply $\xi^\mu\vert_{z=0}$, the restriction of the diffeomorphism on $\partial M$.

One can show that there is a one-to-one correspondence between these asymptotic isometries and conformal transformations of the boundary metric. For example, taking $\hat{\xi}^k=0$, and using the expansion (\ref{expansion}),
\begin{equation}
g_{ij}+\mathcal{L}_{\xi}g_{ij} = \frac{1}{z^2}\left(1-2\hat{\xi}^z\right)\gamma^{(0)}_{ij}+\mathcal{O}(z^0),
\end{equation}
which is nothing but an infinitesimal conformal transformation of the boundary metric $\gamma^{(0)}_{ij}$, also referred to as a ``Penrose-Brown-Henneaux" transformation. As a consequence, an alAdS spacetime has its boundary metric $\gamma^{(0)}$ fixed, up to some conformal transformation or in other words: 
\begin{equation}
    \delta_\xi \gamma^{(0)} \propto \gamma^{(0)}.
\end{equation}
Thus, we may choose either a particular conformal frame in which $\delta_\xi \gamma^{(0)}=0$ or work in a conformally invariant way as above. We will in the following choose the first possibility, but one has to keep in mind the equivalence of the two possibilities, the second being simply a reparametrisation invariance of the variable  $z$.  

The transformation of $\gamma_{ij}$ is given, at leading order, by
\begin{equation}
    \delta_\xi \gamma^{(0)}_{ij}=\mathcal{L}_{\hat{\xi}} \gamma^{(0)}_{ij} - 2\hat{\xi}^z\gamma^{(0)}_{ij}
\end{equation}
Imposing the above to vanish, one sees that $\hat{\xi}^i$ must be a conformal Killing vector of $\gamma^{(0)}$ and that $\hat{\xi}^z = \frac{1}{3}D_k \hat{\xi}^k$ with $D_k$ being the Levi-Civita connection on $(\mathcal{I},\gamma^{(0)})$.  

The expansion (\ref{expansionvierbeins}) should also be kept fixed under asymptotic transformations of the form (\ref{expansion}). We thus require 
\begin{equation} \label{vartetcon}
\delta e^3 = 0, \qquad \delta e_z^{\hat{a}}=0, \qquad \delta e^{(-1)\hat{a}}_k=0.
\end{equation}
One could also require higher order conditions but they won't be needed here. 

The transformation of the vierbeins involves also an internal Lorentz transformation
\begin{equation}
\delta_\Lambda e^a = \Lambda^a{}_{b} e^b.
\end{equation}
Hence, the combined action of the diffeomorphisms and internal Lorentz transformations is given by 
\begin{equation}\label{vartetrads}
    \delta_{\xi,\Lambda} e^a = K_{\xi,\Lambda} e^a= \mathcal{L}_\xi e^a + \Lambda^a{}_b e^b.
\end{equation}
Following \cite{Godazgar:2022foc}, we fix the internal Lorentz transformation so that the combined action as given in equation \eqref{vartetrads} produces the required transformation of the vierbein components set out in \eqref{vartetcon}.  Note that in the case where the diffeomorphism corresponds to a Killing isometry of the background, then the Lie derivative of the tetrad will vanish, which means that a compensating transformation will not be required.  

Assuming an expansion of $\Lambda$ in $z$ of the form
\begin{equation} \label{Lexp}
    \Lambda^{ab}= \Lambda^{(0)ab} + z\Lambda^{(1)ab} + z^2\Lambda^{(2)ab} \dots,
\end{equation}
the conditions given in \eqref{vartetcon} fix the form of $\Lambda$ order by order. 

Consider first  the condition $\delta e^3_z=0$:
\begin{align*}
\delta_{\xi, \Lambda} e^3_z &= \mathcal{L}_\xi e^3_z + \Lambda^3{}_a e^a_z \notag \\
											&= \xi^z \partial_z e^3_z + e^3_z \partial_z \xi^z + \Lambda^3{}_3 e^3_z \notag \\
											&= \partial_z (\xi^z  e^3_z) = \partial_z (\ell \hat{\xi}^z ) =0,
\end{align*}
where we have used the form of the vierbeins \eqref{expansionvierbeins} and the fact that $\Lambda$ is antisymmetric.  Thus, this condition does not impose any constraints on $\Lambda$ and is trivially satisfied.  Similarly, it can be shown that the conditions $\delta e^3_k=0, \ \delta_\xi e^{\hat{a}}_k =0$ imply that
\begin{equation}
\Lambda^{3\hat{a}} = -\ell \partial_k \hat{\xi}^z e^{k\hat{a}}
\end{equation}
and the condition $\delta_\xi e^{(-1)\hat{a}}_k = 0$ implies at leading order that
\begin{align}
 \Lambda^{(0)\hat{a}\hat{b}}=e^{(-1) k [\hat{a}}\mathcal{L}_{\hat{\xi}} e^{(-1)\, \hat{b}]}_k = e^{(-1) i [\hat{a}} e^{(-1) \hat{b}]}_{j} D_i \hat{\xi}^j
\end{align}
where $D_i$ denotes covariant derivative associated with the metric $\gamma^{(0)}_{ij}$.  Since we are only interested in the charges at the boundary, it turns out that we will not need in this work higher order terms in $\Lambda$.  However, for completeness, one can show that
\begin{equation}
 \Lambda^{(1)\hat{a}\hat{b}} =0
\end{equation}
and
\begin{equation}
 \Lambda^{(2)\hat{a}\hat{b}} = e^{(-1) i [\hat{a}}\mathcal{L}_{\hat{\xi}} e^{(1)\, \hat{b}]}_i + \frac{1}{3}e^{(-1) i [\hat{a}} e^{(1) \hat{b}]}_{i} D_j \hat{\xi}^j 
 + e^{(-1) i [\hat{a}} e^{(-1) \hat{b}] j} e^{(-1)\hat{c} k} e^{(1)}_{\hat{c} i} D_{[j} \hat{\xi}_{k]}.
\end{equation}

Moreover, the variation of the on-shell spin connections is of the form\footnote{We simply write $\Lambda^{ab}$ for the Lorentz transformation, but we should keep in mind that it is determined by the choice of $e$ and depends on the diffeomorphism $\xi$ as explained above.}
\begin{equation}\label{varomega}
    \delta_{\xi,\Lambda} \omega^{ab}  = K_{\xi,\Lambda} \omega^{ab}= \mathcal{L}_\xi \omega^{ab} - D\Lambda^{ab},
\end{equation}
where $D$ is the covariant derivative with respect to $\omega$, in other words: 
\begin{equation}
    D\Lambda^{ab} = d\Lambda^{ab} + \omega^{ac}\Lambda_c{}^d - \Lambda^{ac}\omega_c{}^d.
\end{equation}

\section{Asymptotic charges of the AdS theory} \label{sec:charges}

As explained in the introduction, there are several different methods for computing (asymptotic) charges for general gravitational systems.  Here, we follow the covariant phase space formalism; see \cite{Compere:2018aar, Godazgar:2020kqd} for details, which is summarised by the following steps:
\begin{enumerate}
    \item One computes the boundary term, or presymplectic potential $\theta(\phi,\delta \phi)$ from an arbitrary variation of the action:
    \begin{equation}
    \delta L(\phi) = E(\phi) \delta \phi + d\theta(\phi,\delta \phi),
    \end{equation}
    where $\phi$ represents the fields of the theory, $L$ is the Lagrangian and $E$ are the equations of motion.
    \item The exterior derivative of the presymplectic potential on phase space gives the presymplectic 2-form:\footnote{We assume throughout that the variations commute.  In particular, the variation of transformation generators is always taken to be trivial.  Of course, one may consider different slicings of phase space \cite{Adami:2021nnf}.}
    \begin{equation}
    \omega(\phi, \delta_1 \phi, \delta_2 \phi) = \delta_1 \theta(\phi,\delta_2 \phi) - \delta_2 \theta(\phi,\delta_1 \phi).
    \end{equation}
    \item Finally, the variation of the Hamiltonian $\dd H_\xi$ is given by the integral on $\Sigma$ of the presymplectic potential contracted in phase space with a direction associated with the symmetry of interest:
     \begin{equation} \label{Ham}
    \dd H_\tau = \int_\Sigma \omega(\phi, \delta \phi, \delta_\tau \phi).
    \end{equation}
   We write $\dd H_\tau$ since the integral may not be necessarily integrable.  
\end{enumerate}

In general, we would expect (or hope) to transform the volume integral in \eqref{Ham} into a boundary integral over $\partial \Sigma$ so that the charge/Hamiltonian\footnote{We use the terms charge and Hamiltonian interchangeably here.  No confusion should arise from this.} really does only depend on the asymptotic form of the fields.  However, this is done on a case-by-case basis.  For a diffeomorphism generated by $\xi$, this can be done and the variation of the asymptotic charge is given by: 
\begin{equation}
\dd H_\xi= \int_{\partial \Sigma} \Big\{ \delta Q_{\xi} - \iota_{\xi} \theta(\phi,\delta \phi) \Big\},
\end{equation}
where, in the AdS case that we are interested here, $\partial \Sigma$ is a section of $\mathcal{I}$, the conformal boundary, and $Q_{\xi}$ is the Noether charge obtained from the Noether current,
\begin{equation} \label{Ncurr}
j_\xi = dQ_{\xi} = \theta(\phi,\delta_\xi\phi) - \iota_{\xi} L(\phi).
\end{equation}

Thus, as a first step we calculate the presymplectic potential $\theta$ associated with the theory defined by action \eqref{action}.  Varying the associated Lagrangian gives
\begin{equation} \label{vL}
\delta L = 2\kappa P_{abcd}\left( \bar{\mathcal{R}}^{ab}\wedge e^c \wedge \delta e^d - De^a \wedge e^b\wedge \delta \omega^{cd}\right) + d(\kappa\ell^2 P_{abcd} \delta \omega^{ab}\wedge \bar{\mathcal{R}}^{cd}),
\end{equation}
where we make free use of the differential Bianchi identity \eqref{Bianchib} and the Schouten identity. We recognize in the first two terms the equations of motion corresponding to the fields $e$ and $\omega,$ respectively\footnote{Equation $\varepsilon_{abcd}\, T^a \wedge e^b =0$ is equivalent to $T^a =0.$}
\begin{equation}\label{EOM}
\varepsilon_{abcd}\, \bar{\mathcal{R}}^{ab}\wedge e^c=0, \qquad \varepsilon_{abcd}\, T^a \wedge e^b =0.
\end{equation}
 We can read off the presymplectic potential from the third term of \eqref{vL},
\begin{equation}\label{presym}
    \theta(\omega,\delta \omega, e) = \kappa\ell^2 P_{abcd}\, \delta \omega^{ab}\wedge \bar{\mathcal{R}}^{cd}.
\end{equation}
The Noether current, as defined in \eqref{Ncurr}, is then of the form
\begin{align}
j_\xi &=  \kappa\ell^2 P_{abcd}\, \Big( K_{\xi,\Lambda} \omega^{ab} - \iota_\xi \bar{\mathcal{R}}^{ab}  \Big) \wedge \bar{\mathcal{R}}^{cd} \notag \\
&=  \kappa\ell^2 P_{abcd}\, \Big( (\mathcal{L}_{\xi}- \iota_\xi d) \omega^{ab} - d\Lambda^{ab} + [\iota_\xi \omega-\Lambda, \omega]^{ab}  \Big) \wedge \bar{\mathcal{R}}^{cd} \notag \\
&= d \Big(\kappa\ell^2 P_{abcd}\, ( \iota_\xi \omega - \Lambda )^{ab} \, \bar{\mathcal{R}}^{cd} \Big)  + \kappa\ell^2 P_{abcd}\, \Big(  [\iota_\xi \omega - \Lambda, \omega]^{ab}  \wedge \bar{\mathcal{R}}^{cd} - (\iota_\xi \omega - \Lambda)^{ab} [\bar{\mathcal{R}}, \omega]^{cd} \Big) 
       \notag \\
&= d \Big(\kappa\ell^2 P_{abcd}\, ( \iota_\xi \omega - \Lambda )^{ab} \, \bar{\mathcal{R}}^{cd} \Big),
\end{align}
where in the second equality, we have used equation (\ref{varomega}), \eqref{curvature} and \eqref{EOM}, in the third equality, we have used the Cartan magic formula
\begin{equation}
\mathcal{L}_{\xi} = \iota_\xi d + d \iota_\xi
\end{equation}
and Bianchi identity \eqref{Bianchib} and in the final equality we have used the Schouten identity.  Therefore, we conclude that the Lorentz invariant Noether charge is
\begin{equation}\label{two-form}
Q_{\xi} = \kappa\ell^2  P_{abcd}\left( \iota_{\xi} \omega^{ab} - \Lambda^{ab}\right)\bar{\mathcal{R}}^{cd}.
\end{equation}

Now that we have an expression for the Noether charge, we can write down an expression for $\dd_{\xi} H$\footnote{Note that we treat $\Lambda$ as a transformation parameter and therefore set its arbitrary variation to zero.  This is consistent with the order by order expansion of $\Lambda$ discussed in section \ref{sec:symm}.}
\begin{align}
    \dd_{\xi} H = \int_{\partial \Sigma} \delta Q_{\xi, \Lambda} &- \iota_{\xi} \theta(e,\omega, \delta \omega)\\
    =&\kappa \ell^2 P_{abcd}\int_{\partial \Sigma} \left( \iota_{\xi} \omega^{ab} - \Lambda^{ab}\right)\delta \bar{\mathcal{R}}^{cd} + \delta \omega^{ab}\wedge \iota_{\xi} \bar{\mathcal{R}}^{cd} \notag \\
    =&\kappa \ell^2  P_{abcd}\int_{\partial \Sigma} \left(\iota_{\xi} \omega^{ab} - \Lambda^{ab}\right) \delta \mathcal{R}^{cd} + \delta \omega^{ab}\wedge \iota_{\xi} \mathcal{R}^{cd} \notag \\
    &+ 2\kappa P_{abcd} \int_{\partial \Sigma} \left(\iota_\xi \omega^{ab} - \Lambda^{ab}\right) \delta e^c\wedge e^d+ \delta \omega^{ab}\iota_\xi e^c\wedge e^d.
\end{align}
The last line is the variation of the charge in the flat case: 
\begin{equation}
\dd H_\xi^{flat} = 2 \kappa P_{abcd} \int_{\partial \Sigma} \left(\left(\iota_{\xi} \omega^{ab}-\Lambda^{ab}\right) \delta e^c + \iota_\xi e^c \delta \omega^{ab}\right) \wedge e^d,
\end{equation}
while the remaining terms are contributions that exist because of the non-zero cosmological constant.  These terms can be simplified to a single expression \cite{Godazgar:2020gqd}
\begin{align} \label{intparts}
\kappa \ell^2  P_{abcd}\int_{\partial \Sigma} \left(\iota_{\xi} \omega^{ab} - \Lambda^{ab}\right) \delta \mathcal{R}^{cd} + \delta \omega^{ab}\wedge \iota_{\xi} \mathcal{R}^{cd}
=\kappa \ell^2 P_{abcd} \int_{\partial \Sigma}\delta \omega^{ab} \wedge K_{\xi,\Lambda} \omega^{cd}.
\end{align}

In summary, the variation of the asymptotic charge is 
\begin{equation}\label{charges}
    \dd_{\xi} H = \dd H_{\xi,1} + \dd H_{\xi,2} + \dd H_{\xi,3}, 
\end{equation}
where
\begin{align}
    &\dd H_{\xi,1} = \kappa \ell^{2}  P_{abcd}\int_{\partial \Sigma} \delta \omega^{ab} \wedge K_{\xi,\Lambda} \omega^{cd}, \label{H1}\\
    & \dd H_{\xi,2} = 2\kappa P_{abcd} \int_{\partial \Sigma} \left(\iota_{\xi} \omega^{ab}- \Lambda^{ab}\right) \delta e^c\wedge e^d, \label{H2} \\
    & \dd H_{\xi,3} = 2\kappa P_{abcd} \int_{\partial \Sigma} \iota_{\xi} e^c\, \delta \omega^{ab}\wedge e^d.  \label{H3}
\end{align}
Of course, there is no reason to believe that (\ref{H2}) and (\ref{H3}) should be decoupled and in fact, they are not. We split the charge in the same way as in the flat case \cite{Godazgar:2020kqd}, with \eqref{H1} corresponding to the higher-derivative charges studied in \cite{Godazgar:2022foc}.

\section{Asymptotic charges of alAdS spacetimes} \label{sec:alAdScharges}

Now that we have the general expression for the asymptotic diffeomorphism charges of the AdS theory, we need to evaluate the expressions for the symmetry generators derived in section \ref{sec:symm} and the variation of the fields.  Recall that the only non-vanishing variations is the traceless part of $\gamma^{(3)}$ and the corresponding vierbein component $e^{(2)\hat{a}}_i$.  Thus, we keep $\delta \gamma^{(3)}$ non-zero while keeping in mind that $\gamma^{(0)ij}\delta \gamma^{(3)}_{ij}=0$. 

First, let us study the asymptotics of the spin connections.  We will find that this will dramatically simplify the computations. On-shell, the spin connections are given by the torsion-free condition
\begin{equation}
    \omega_{ab\, \mu} = e^\nu_{\left[a\right.} \partial_\mu e_{\left.b\right]\nu}+e^{\nu}_{\left[a\right.}e^\sigma_{\left.b\right]} \partial_\sigma g_{\mu\nu}.
\end{equation}

From the antisymmetry of $\omega^{ab}$ and $g_{z\hat{a}}$=0, it can be easily observed that 
\begin{equation} \label{spinz}
\omega^{3\hat{b}}_z = 0, \qquad \omega^{\hat{a}\hat{b}}_z=\mathcal{O}(z), \qquad \delta \omega^{\hat{a}\hat{b}}_z=\mathcal{O}(z^2).
\end{equation}
We will find later that these components do not contribute to the charges.  The other components are of the form
\begin{align}
    \omega^{3\hat{b}}_k &= \frac{1}{\ell z}e^{(-1)\hat{b}}_k + \frac{z}{\ell} \left(e^{(1) \hat{b}}_k - \gamma^{(2)}_{kl} e^{(1) \hat{b} l}\right) + \frac{z^2}{\ell} \left(e^{(2)\hat{b}}_k -\frac{3}{2}\gamma^{(3)}_{kl} e^{(-1)\hat{b} l} \right)+\mathcal{O}(z^3) \notag \\
    \delta\omega^{3\hat{b}}_k &= \frac{z^2}{\ell} \left(\delta e^{(2)\hat{b}}_k -\frac{3}{2}\delta \gamma^{(3)}_{kl} e^{(-1)\hat{b} l} \right)+\mathcal{O}(z^3)
    \label{expanspinco1}
\end{align}
and
\begin{equation}
        \omega^{\hat{a}\hat{b}}_k =  \omega^{(0)\hat{a}\hat{b}}_k +\mathcal{O}(z), \qquad  \delta \omega^{\hat{a}\hat{b}}_k =\mathcal{O}(z^3),
    \label{expanspinco2}
\end{equation}
where $\omega^{(0)\hat{a}\hat{b}}_k$ corresponds to the spin connection associated with dreibein $e^{(-1)\hat{a}},$ i.e.\ it corresponds, at leading order, to the spin connection of the 3-manifold $(\mathcal{I},\gamma^{(0)})$ and is thus fixed by the equations of motion at leading order.

From the fact that we find that $\delta \omega^{ab} = \mathcal{O}(z^2),$ we conclude simply that
\begin{equation}
\dd H_{\xi,1} = \mathcal{O}(z^2), 
\end{equation}
hence, it does not contribute to the boundary charges of the alAdS background.  We are then left with $\dd H_{\xi,2}$ and $\dd H_{\xi,3}$.  Moreover, from the expansion of the Lorentz parameter \eqref{Lexp} and the fact that $\delta e^a = \mathcal{O}(z^2)$, we find that the second term in $\dd H_{\xi,2}$ vanishes, so that
\begin{equation}
\dd H_{\xi} = 2\kappa P_{abcd} \int_{\partial \Sigma} (\iota_{\xi} \omega^{ab}  \delta e^c+ \iota_{\xi} e^c\, \delta \omega^{ab} ) \wedge e^d + \mathcal{O}(z).
\end{equation}

In analogy with the flat case \cite{Godazgar:2020kqd}, we choose to split the analysis into an electric and magnetic/dual part.  The electric contribution is given by the $\varepsilon$ part of $P_{abcd},$ as defined in \eqref{P}, while the magnetic/dual part is controlled by the parameter $\lambda.$

\subsection{Electric charges}
Using the expansion (\ref{expanspinco1}), the fact that the boundary $\mathcal{I}$ does not extend along the direction $z$ and that at least one of the indices has to be $3$, it is simple to show that
\begin{align}
    \dd \mathcal{Q}_{\xi}|_{z=0}  &=2\kappa\, \varepsilon_{3\hat{b} \hat{c} \hat{d}} \int_{\partial \Sigma} (\iota_{\xi} \omega^{3\hat{b}}  \delta e^{\hat{c}}+ \iota_{\xi} e^{\hat{c}}\, \delta \omega^{3 \hat{b}} ) \wedge e^{\hat{d}} \notag \\
    &= \frac{3\kappa}{\ell}\, \varepsilon_{3\hat{b}\hat{c}\hat{d}}\int_{\partial\Sigma} \hat{\xi}^k e^{(-1)\hat{b}}_k\, e^{(-1)\hat{c}l}\,  e^{(-1)\hat{d}}_J \, \delta \gamma^{(3)}_{lI}  dx^I \wedge dx^J \notag \\
    &=-\frac{3\kappa}{2\ell}\, \varepsilon_{3\hat{b}\hat{c}\hat{d}}\int_{\partial\Sigma} \hat{\xi}^k \delta \gamma^{(3)}_{kl} e^{(-1)\hat{b} l}\ e^{(-1)\hat{c}}\wedge e^{(-1)\hat{d}},
\end{align}
where $\lbrace x^I\rbrace$ are coordinates on $\partial\Sigma$ as a section of $\mathcal{I}$.  Note that we have used a Schouten identity in the final equality and the trace-free condition on $\gamma^{(3)}$.

Thus, we have an integrable charge, that coincides with the  usual Brown-York charge
\begin{equation} \label{BY}
    Q_{BY}\left[\xi\right] = -\frac{3\kappa}{\ell} \int_{\partial
    \Sigma} \sqrt{q}\, \hat{\xi}^k \gamma^{(3)}_{kl} n^l d^2x,
\end{equation}
where $n$ is an outward pointing normal vector to $\partial \Sigma$ such that $\gamma^{(0)}(n,n)=-1$ and $\sqrt{q} d^2x$ is the area element on $\partial\Sigma$. 

Consider the Schwarzschild-AdS solution, which has a Fefferman-Graham form, with the coordinates on the boundary, the usual $(t,\theta,\phi)$ coordinates:\footnote{The Fefferman-Graham form of the metric is obtained from that in the usual Schwarzschild coordinates by setting ${r(z) = \frac{1}{z}+\frac{3}{4\Lambda}z-\frac{m}{\Lambda}z^2+\mathcal{O}(z^3)}$.}
\begin{gather}
        \gamma^{(0)}(x^i) = -\frac{1}{\ell^2}dt^2 + d\Omega_2^2, \qquad 
        \gamma^{(2)}(x^i) = -\frac{1}{2}\left(dt^2 + \ell^2d\Omega_2^2\right) \notag \\
        \gamma^{(3)}(x^i) = \frac{4}{3} m G_N dt^2 +6\ell^2m G_N d\Omega_2^2.
\end{gather}
Thus,  the section $\partial\Sigma$ on $\mathcal{I}$ is simply $S^2$ and the normal $n$ is $\ell\partial_t$.  Plugging these expressions into equation \eqref{BY} with $\hat{\xi}$ chosen as $-\partial_t$ and $\partial_{\phi}$, one recovers the expected energy and angular momentum,
\begin{equation}
    E = Q_{BY}\left[-\partial_t\right] = m, \quad L = Q_{BY}\left[\partial_\phi\right] = 0.
\end{equation}
\subsection{Dual charges}

The dual charges involve contractions of each term in the charges. Since only $\delta e^{\hat{a},(2)}_\mu$ varies, one gets for $\dd H_{\xi,2}$: 
\begin{align}
    \dd \widetilde{\mathcal{Q}}_{\xi}|_{z=0}   &= 2 \kappa  \int_{\partial \Sigma} (\iota_{\xi} \omega_{ab}  \delta e^a+ \iota_{\xi} e^a\, \delta \omega_{ab} ) \wedge e^b \notag \\
    &= 2 \kappa \int_{\partial \Sigma} (\iota_{\xi} \omega_{\hat{a}\hat{b}}  \delta e^{\hat{a}}+ \iota_{\xi} e^a\, \delta \omega_{a\hat{b}} ) \wedge e^{\hat{b}}.
\end{align}
From the expressions for vierbein and the spin connection, \eqref{expansionvierbeins}, \eqref{expanspinco1} and \eqref{expanspinco2}, it is simple to see that
\begin{equation*}
 \iota_{\xi} \omega_{\hat{a}\hat{b}}  \delta e^{\hat{a}}+ \iota_{\xi} e^a\, \delta \omega_{a\hat{b}} = \mathcal{O}(z^2),
\end{equation*}
which implies that the dual charge
\begin{equation}
\dd \widetilde{\mathcal{Q}}_{\xi} = \mathcal{O}(z).
\end{equation}
As a consequence, the dual charges identically vanish at the boundary of alAdS spacetimes with a Dirichlet boundary condition.  This is in contrast to the flat case, where dual charges do contribute at leading order at null infinity \cite{Godazgar:2020kqd}.

\section{Taub-NUT-AdS} \label{sec:NUT}

We have seen in the previous section that dual charges are trivial on the boundary of alAdS backgrounds satisfying the Dirichlet condition. We consider in this section an example of an alAdS black hole solution giving non-trivial dual charges at the boundary.

\subsection{Preliminaries}

The Taub-NUT-AdS spacetimes are a 2-parameter family of metrics, parametrised by a mass parameter $m$ and NUT parameter $n$,  which in local Boyer-Lindquist-like coordinates\footnote{$r_+$ corresponds to the horizon radius given by the maximal root of $\mathcal{Q}(r)=0$. In the flat case, one has simply $r_+ = m + \sqrt{m^2+n^2}$. } $(t,r,\theta,\phi)\in \mathbb{R}\times \left(r_+,\infty\right)\times \left(0,\pi\right)\times \left[0,2\pi\right)$ take the form \cite{Griffiths:2009dfa}
\begin{equation} \label{NUTads}
    ds^2 = -\frac{Q}{\Sigma}\left( dt - A\right)^2 + \frac{\Sigma}{Q} dr^2 + \Sigma\left( d\theta^2 +\sin^2\theta d\phi^2\right)
\end{equation}
with\footnote{For brevity, we have set $G_N=1$.}
\begin{gather}
Q= r^4\ell^{-2} +r^2\left( 1+6n^2\ell^{-2}\right) -2mr-3n^4\ell^{-2} -n^2 , \qquad \Sigma = n^2+r^2, \notag \\
 A = -2n\left( \cos\theta + \sigma\right) d\phi.
\end{gather}

This vacuum solution exhibits some interesting features: 
\begin{itemize}
    \item The NUT parameter $n$, can be seen in many ways as a monopole charge, the solutions being thus analogous to the Dirac monopole \cite{MahdiTaubNUT}. It may also be viewed as an angular momentum source. 
    \item Just like the Dirac monopole, there exists a wire singularity.  The location of this singularity on the sphere can be moved by adjusting the constant $\sigma$, which represents the freedom to shift $t$: $t\rightarrow t+ \textup{constant} \times \phi$.  $\sigma = \pm 1$ corresponds to having the singularity at the south and north pole, respectively, while  $\sigma =0$ distributes the singularity symmetrically between the two poles.
    \item The solution possesses closed timelike and null geodesics for $\abs{\sigma} >1$. For this reason, we will only consider the case $\abs{\sigma}\leq 1$. 
\end{itemize}

One way of dealing with the wire singularity, suggested by Misner \cite{Misner63} is to consider the Euclidean metric with the identification of $t$ such as: $t\sim t+8\pi n$. However, this constrains the parameters of the solution, implying that
\begin{equation}\label{selfdual}
    m = n\left( 1+\frac{4n^2}{\ell^2}\right),
\end{equation}
which makes it difficult to analyse the thermodynamics of the solution. 

We shall stick with the Lorentzian solution and make no assumption on the metric, except on the value of $\sigma$ outlined above. As argued in \cite{Hennigar2019}, this spacetime is in fact less pathological than is usually claimed. We shall find that one recovers the expected conserved charges and that the first law can be written using the full alAdS action and its corresponding charges.

Metric \eqref{NUTads} can be put in a Fefferman-Graham form \eqref{FGexpansion1} using the expansion
\begin{equation}
    r(z) = \frac{1}{z}-\frac{1}{4} z \left(5 n^2+\ell^2\right)+\frac{\ell^2}{3}m z^2+n^2 z^3 \left(n^2+\frac{\ell^2}{4}\right)+ \mathcal{O}(z^3)
\end{equation}
with $z\in \left(0,z_0\right)$ an inverse radius coordinate such that $r(z_0)=r_+$, giving
\begin{equation}
    \begin{cases}
        \gamma^{(0)} = -\frac{1}{\ell^2}(dt-A)^2 + d\Omega_{2}^2 \\
        \gamma^{(2)} = -\frac{1}{2}\left(1+5\frac{n^2}{\ell^2} \right)\left(dt-A \right)^2 + \frac{1}{2}\left( -\ell^2+3n^2 \right)d\Omega^2_{2}\\
        \gamma^{(3)}=\frac{4}{3}m\left(dt-A \right)^2 +6\ell^2md\Omega_2^2
    \end{cases}
\end{equation}
Observe that $\gamma^{(0)}$ is not the metric on the cylinder $\mathbb{R}\times S^2$: the NUT parameter $n$ modifies the global topology of the boundary.  Moreover, as a consequence  $\delta\gamma^{(0)},\delta\gamma^{(2)},\delta \gamma^{(3)}\neq 0$.  Therefore, the analysis of section \ref{sec:alAdScharges} no longer applies.

We choose a canonical frame $e^a$ with 
\begin{equation} \label{TNbasis}
      e^0 = - \sqrt{\frac{Q}{\Sigma}}\left( dt -A\right), \qquad  e^1 = \sqrt{\Sigma}d\theta,  \qquad e^2 = \sqrt{\Sigma}\sin{\theta} d\phi, \qquad 
        e^3 = \ell z^{-1}dz,
\end{equation}
whose expansion in terms of $z$ is of the form
\begin{align}
    e^0 &= -\frac{1}{z}\left( 1 + \frac{1}{4}(\ell^2+5{n^2})z^2 + \frac{2}{3}mz^3 +\mathcal{O}(z^4)\right)(dt - A), \notag \\
    e^1 &= \frac{1}{z}\left( 1 - \frac{1}{4}(3n^2+\ell^2)z^2 + \frac{\ell^2}{3}mz^3 + \mathcal{O}(z^4)\right)d\theta, \notag \\
    e^2 &= \frac{1}{z}\left( 1 - \frac{1}{4}(3n^2+\ell^2)z^2 + \frac{\ell^2}{3}mz^3+ \mathcal{O}(z^4)\right)\sin{\theta}d\phi.
\end{align}
As defined in \eqref{expansionvierbeins}, we will refer to $e^{(k)\hat{a}}$ with $k=-1,1,2$ for the first, second and third term in the expansions above, respectively. 

The goal is now to compute the charges and dual charges for the Taub-NUT-AdS solution. The expectation is that the appearance of $n$ as a parameter will give a non-trivial contribution to the dual charges. 

\subsection{Charges of Taub-NUT-AdS}

In this section we shall compute the electric and dual charges corresponding to the symmetry generators $\xi^\mu$ given in equation (\ref{expansion}).  Note that the charges derived in section \ref{sec:charges}, namely expressions (\ref{H1}),(\ref{H2}),(\ref{H3}) must be supplemented in this case by a total derivative term that we dropped there.  Indeed, expression (\ref{H1}) was obtained in equation (\ref{intparts}) using an integration by parts and discarding the boundary terms. Those terms, however, cannot be simply thrown away for Taub-NUT backgrounds because the presence of a wire singularity makes the boundary non-smooth at $\theta = \lbrace0,\pi\rbrace$ depending on the value of $\sigma$. Thus, we must include this non-trivial boundary term, which is of the form
\begin{equation} \label{bdr}
    \dd \psi_{\xi} + i\lambda \dd \widetilde{\psi}_{\xi} = \kappa \ell^2 P_{abcd}\int_{\partial\Sigma} d\left( \iota_\xi \omega^{ab}\delta \omega^{cd}\right).
\end{equation}
As we shall see below, this term actually plays an important role in the regularisation of the charges and behaves like a counter-term. Since this term does not appear in the Schwarzschild-AdS case, it should naturally only depend on $\delta n$.  

In order to derive the charges, we substitute the  vierbeins \eqref{TNbasis} into the expression for the spin connections given in equations \eqref{spinz}, \eqref{expanspinco1} and \eqref{expanspinco2}.  It is simple to see that $\dd \widetilde{H}_{1,\xi}$ will still not contribute.  

The full dual charge $\dd \widetilde{\mathcal{Q}}_\xi = \dd \widetilde{H}_{2,\xi} + \dd \widetilde{H}_{3,\xi} + \dd\widetilde{\psi}$ at the boundary is given by
\begin{align}
    \delta \widetilde{\mathcal{Q}}_{\xi}|_{z=0}= -4\kappa\int_{\partial \Sigma} &\left\{\delta \left[ n \left(1+\frac{4 n^2}{\ell^2} \right)\right] \hat{\xi}^t(t,\theta,\phi)\right.\\
    &\left.+3\delta \left[n^2 \left(1+\frac{4 n^2}{\ell^2} \right)\right] (\cos \theta+\sigma) \hat{\xi}^\phi(t,\theta,\phi)\right\}  \sin (\theta),
\end{align}
while the electric charge is 
\begin{align}
    \delta \mathcal{Q}_\xi|_{z=0} = -4 \kappa \int_{\partial\Sigma} \left(3 (\cos\theta+\sigma) \delta( mn ) \, \hat{\xi}^\phi(t,\theta,\phi)+ \delta m\, \hat{\xi}^t(t,\theta,\phi))\right)\sin\theta.
\end{align}

In deriving these charges, it is interesting to note the role of the boundary term \eqref{bdr}.  It is in fact a singular term that is of order $\mathcal{O}(z^{-1})$.  Therefore, it acts as a counter-term cancelling divergent terms in the other contributions to the charges and ensuring that the final result is finite and perfectly well-defined at the boundary. 

For $\hat{\xi}^\mu$ constant, one expects to reproduce the usual energy $\delta E \sim \delta m$ and the dual energy (interpreted as a magnetic charge) $\delta \widetilde{E} \sim \delta n$. This is indeed the case, although one also has angular momentum contributions and higher order terms:\footnote{We have used the fact that we have set $G_N=1$ and so $16 \pi \kappa =1.$}
\begin{align*}
    &\delta \widetilde{\mathcal{Q}}_{\xi}=-  \hat{\xi}^t\delta \left( n+\frac{4 n^3}{\ell^2}\right) - 3\hat{\xi}^\phi\sigma\delta \left(  n^2+\frac{4 n^4}{\ell^2} \right) \\
    & \delta \mathcal{Q}_\xi =-\hat{\xi}^t \delta m-3 \hat{\xi}^\phi  \sigma \delta (mn)
\end{align*}
Thus,
\begin{gather}
        E = m, \qquad L = -3\sigma mn=-3\sigma nE \notag \\[2mm]
        \widetilde{E} = n\left( 1+\frac{4 n^2}{\ell^2}\right),\qquad \widetilde{L} = -3n^2\left(1+\frac{4 n^2}{\ell^2}\right)=-3\sigma n\widetilde{E}
        \label{Charges}
\end{gather}
Observe how the flat limit gives the expected dual energy and angular momentum. Note also that the NUT parameter appears in both the dual energy and the angular momentum. Recall that $n$ is usually interpreted as either an angular momentum parameter or as a magnetic charge. Briefly, we review the  justifications for these two points of view.

\paragraph{The NUT parameter $n$ as an angular momentum:}
    The rotation parameter $\Omega$ can be read off from the metric components to be: 
    \begin{equation}
        \Omega = -\frac{g_{t\phi}}{g_{\phi\phi}} = \frac{\mathcal{Q}^2A_\phi}{-A_\phi^2\mathcal{Q}^2+\Sigma^3\sin^2\theta}
    \end{equation}
    Such a rotation vanishes at the horizon, $\Omega_\mathcal{H}=0$, but turns out to be non-zero at the boundary since \[\Omega_\infty=\frac{1}{2n\left(\sigma+\cos\theta\right)} \neq 0.\] In the $\sigma =0$ case, one has two Misner strings rotating in opposite directions since\footnote{See \cite{Manko:2005nm} for more details.} 
    \begin{equation}
        \left. \Omega \right|_{\theta=0, \sigma =0} = \frac{1}{2n} \quad \quad  \left. \Omega \right|_{\theta=\pi, \sigma =0}=-\frac{1}{2n}
    \end{equation}
The vanishing of $L$ and $\widetilde{L}$ in this case simply can be understood in terms of the cancelling of the opposite angular momenta generated by the Misner string. Observe the similarity with the Kerr solution since in that case $L_{Kerr}=ma$ with $a$ the angular momentum parameter. The factor of $3$ in the expression for $L$ is because the natural Killing vector to consider is $\chi = \partial_\phi - 2\sigma n \partial_t$ rather than $\partial_\phi$ and in this case: 
    \begin{equation}
       L_\chi = -\sigma m n
    \end{equation}
    
\paragraph{The NUT parameter $n$ as a magnetic charge:}
Viewing $A=-2n\left(\cos\theta + \sigma\right)d\phi$ as a gauge form, the different fluxes are given by
    \begin{equation}
       \frac{1}{8\pi}\int_{\partial\Sigma} \star dA =0, \quad \quad \frac{1}{8\pi} \int_{\partial\Sigma} dA  = n.
    \end{equation}
In fact, in the asymptotically flat case, taking a characteristic initial value problem point of view, one can make this statement more precise by showing how one may construct the Taub-NUT solution from a Dirac monopole solution \cite{MahdiTaubNUT}.  The relation is not as simple in the AdS case since the dual energy given by the Holst term is \footnote{Note that the solution suggested by Misner (\ref{selfdual}) is simply a self-dual statement.} $\widetilde{E} = n\left(1-\frac{4}{3}\Lambda n^2\right)$. 

With the first law of black hole mechanics in mind, rather than viewing the NUT parameter solely as a magnetic charge or solely as an angular momentum parameter, it may be more fruitful to view it as playing both of these roles \cite{Durka:2019ajz}.

\subsection{Black Hole First Law}

We have found the conserved quantities for the general Taub-NUT solution: $E,J_\sigma, \Tilde{E},\Tilde{J}_\sigma$ with $J_\sigma,\Tilde{J}_\sigma$ vanishing if $\sigma=0$.

One would like to relate the charges we have just computed to thermodynamic quantities defined on the horizon of the black hole. To do this, we need to take into account the full topology of the spacetime. The Cauchy surface $\Sigma$ has as a boundary a combination of the sphere at infinity $S_\infty$, where we computed the charges, the horizon $\mathcal{H}$ and finally the two Misner strings, where the metric becomes singular, $\mathcal{S}_{\pm}$. The boundary of the Cauchy surface can thus be taken to be: 
\begin{equation}\label{boundary}
    \partial \Sigma = S_\infty -\left( \mathcal{H} +\mathcal{S}_+ - \mathcal{S}_-\right),
\end{equation}
where the minus signs are due to the orientation. 

Let us take $\xi$ to be the Killing null generator of the horizon,  $\xi=\partial_t$. Since $\xi$ is Killing, the integral of the pre-symplectic form $\omega$ on the whole Cauchy surface vanishes identically: 
\begin{equation} \label{om}
    \int_{\Sigma} \omega\left(\psi, \delta \psi, \delta_\xi \psi\right)=0.
\end{equation}
Defining $k_\xi \equiv \delta Q_\xi - \iota_\xi \theta(e,\omega,\delta \omega)$, we then have that
\begin{align}
    0 &= \int_{\partial\Sigma} k_\xi = \int_{S_\infty}k_\xi -\int_{\mathcal{H}}k_\xi + \int_{\mathcal{S}_+}k_\xi - \int_{\mathcal{S}_-}k_\xi.
\end{align}
As observed in \cite{Frodden:2021ces}, a simple computation shows that the integration on the Misner strings split into terms depending only either on $S_\infty$ or on $\mathcal{H}$ in the following way: 
\begin{equation}
    \int_{\mathcal{S}_+}k_\xi = K_\xi^+(S_\infty)-K_\xi^+(\mathcal{H})
\end{equation}
and similarly for $\mathcal{S}_-$. As a consequence, we can rewrite equation \eqref{om} as
\begin{equation}
    \int_{S_\infty} k_\xi +K^+_\xi(S_\infty)-K^-_\xi(S_\infty)= \int_{\mathcal{H}} k_\xi +K^+_\xi(\mathcal{H})-K^-_\xi(\mathcal{H}).
\end{equation}
This relation forms the basis of the first law: it relates global quantities calculated at infinity to thermodynamic quantities defined at the horizon.  Of course, the challenge then is to identify these quantities.  This is what we now turn to.

For simplicity, we set $\Lambda = 0$, since the cosmological constant does not play an important role in this discussion.\footnote{See Appendix \ref{general} for the general expressions with $\Lambda \neq 0$.}  Moreover, we consider the symmetric case $\sigma=0$. We know from \eqref{Charges} that the integral at infinity is simply equal to the total energy:
\begin{equation}
    \delta H_{\partial_t} = -\delta m -i\lambda \delta n.
\end{equation}
The integral on the horizon gives, on the other hand, after using $m=\frac{1}{2r_+}\left( r_+^2-n^2\right)$: 
\begin{equation} \label{hor}
    \int_{\mathcal{H}} k_\xi = -\delta m \,\frac{-i\lambda nr_+ +r_+^2}{(r_+^2+n^2)}-\delta n\, \frac{n^3-2i\lambda n^2r_+ +3nr_+^2}{2r_+(n^2+r_+^2)}
\end{equation}

We find that the contribution of the Misner strings at infinity $K^+_\xi(S_\infty)$ and $K^-_\xi(S_\infty)$ vanish. Moreover, since the integrals are proportional to $\cos\theta$, we have that $K^+_\xi(\mathcal{H})=-K^-_\xi(\mathcal{H})$.  Thus,
\begin{equation}
  K^+_\xi(\mathcal{H})-K^-_\xi(\mathcal{H})= -\delta m \, \frac{n\left( n+i\lambda r_+\right)}{2(n^2+r_+^2)}- \delta n \frac{n^3 +3nr_+^2 -2i\lambda\left( 2n^2r_+ + r_+^3\right)}{4r_+(n^2+r_+^2)}
\end{equation}

Writing the integral on the horizon \eqref{hor} in terms of $\delta r_+$ instead of $\delta m$ allows us to relate it to the usual definition of the black hole entropy as area given by a Noether charge \cite{Wald:1993nt}
\begin{equation} \label{horint}
\int_{\mathcal{H}} k_\xi = \frac{1}{4\pi r_+} \pi \delta \left(n^2+r_+^2 \right)-i\lambda \frac{n}{2r_+}\delta r_+= T_H \frac{\delta A}{4}-i\lambda \frac{n}{2r_+}\delta r_+,
\end{equation}
where
\begin{equation}
    A = \int_\mathcal{H}\sqrt{\gamma} = \int_{S^2}\Sigma\vert_{r=r_+} \sin\theta =4 \pi \left( r_+^2 + n^2\right)
\end{equation}
and
\begin{equation}
    T_H = \frac{f'(r_+)}{4\pi}, \qquad f(r) = Q/\Sigma.
\end{equation}
Similarly, the contribution from the Misner string can be written as: 
\begin{align} \label{stringint}
    K_\xi^+(\mathcal{H}) - K^-_\xi (\mathcal{H}) = \frac{1}{4\pi n} \delta \left(-2\pi\frac{ n^3}{r_+}\right)+i\lambda \left[\frac{n}{2r_+}\delta r_+ +  \delta n \right],
\end{align}
where we recognise in the first term the Misner potential $\psi=\frac{1}{4\pi n}$ and the Misner charge $N=-2\pi \frac{n^3}{r_+}$ \cite{Bordo:2019tyh}. 
And one obtains a first law, on the electric side at least:
\begin{equation}
    \delta m = T_H \delta S_H + \psi \delta N.
\end{equation}
An important point to observe here is that the two thermodynamic quantities $S_H,N$ are not Noether charges. If one indeed computes the Noether charge at the horizon, as prescribed by Wald \cite{Wald:1993nt}, using $\xi=-\frac{1}{T}\partial_t$, one obtains 
    \begin{equation}
        \dd \mathcal{S} = \delta S_H + \frac{\psi}{T_H}\delta N +i\lambda 4\pi r_+ \delta n.
    \end{equation}
This ``complex" entropy $\mathcal{S}$ is clearly non-integrable. Thus, the presence of the Misner strings seems to make the notion of entropy less clear.  In some sense, the black hole entropy has to take into account the presence of the Misner strings.

There is an alternative interpretation for the quantities we have in the electric part. Using the generating Killing vectors of the Misner strings $\xi_\pm = \partial_t \mp \frac{1}{2n}\partial_\phi$, we can define a \emph{Misner entropy}\cite{BallonBordo:2019vrn}:
\begin{equation}
    K_{\xi_\pm}^\pm(\mathcal{H}) =\mp \frac{\kappa}{4\pi} \delta A_M\mp\frac{\delta r_+}{4}
\end{equation}
with $\kappa=\frac{1}{2n}$ the surface gravity of the Misner strings and $\delta A_M = -2\pi\delta\left(nr_+\right)$ the variation of the ``area" of the Misner string, up to some divergent terms. To obtain the full first law, it is also necessary to compute the previous integrals with respect to a rotation vector $\eta_\pm = \pm\frac{1}{2n}\partial_\phi$:
\begin{equation}
    K_{\eta_\pm}^\pm(\mathcal{H})=\Omega_{\pm} \delta J \pm\frac{\delta r_+}{4}\mp i\lambda\left(\delta r_+ \frac{n}{4r_+} +\frac{\delta n}{2}\right),
\end{equation}
with $\Omega_\pm=\pm\frac{1}{2n}$ and $J=mn$, giving the following first law in the electric part:
\begin{align}
    \delta E = \delta m &= \int_\mathcal{H} k_{\partial_t} + K_{\xi_+}^+(\mathcal{H})+K_{\eta_+}^+(\mathcal{H}) -K_{\xi_-}^-(\mathcal{H})-K_{\eta_-}^-(\mathcal{H})\\
    &=T_H\delta S_H + 2\kappa \delta A_M +2\abs{\Omega} \delta J . 
\end{align}
All the thermodynamic functions we computed here ($S_H,A_M,J$) are integrable but, as discussed earlier, cannot be interpreted as Noether charges. 

So far in this discussion, we have ignored the dual contributions.  We now turn to this.  Note from equations \eqref{horint} and \eqref{stringint} that both the black hole and the Misner strings seem to carry non-trivial thermodynamic dual quantities. To be precise, we have:
\begin{equation}
    \delta \widetilde{E}= \delta n = \underbrace{-\frac{n}{2r_+}\delta r_+}_{\text{Horizon}}+ \underbrace{\delta n + \frac{n}{2r_+}\delta r_+}_{\text{Misner}}.
\end{equation}
These quantities ought to correspond, in some sense, to non-trivial thermodynamic quantities. An example would be
\begin{equation}
    \delta \widetilde{E}=-\frac{n}{4\pi r_+^2}\delta(\pi r_+^2) + \frac{1}{2r_+n}\delta(n^2r_+),
\end{equation}
where each of the terms in the expression above would have a particular physical interpretation, which thus far eludes us.  

A perhaps more natural way to treat the first law in the presence of dual charges is to complexify it, thereby obtaining a \emph{complex first law}.  Such an interpretation is suggested by the fact that the Hamiltonian conjugate to time translations is itself complex, with the real part corresponding to the mass and the imaginary part corresponding to the NUT charge.  Of course, a specific value of $\lambda$ must be chosen in this case to obtain a truly complex law.  Previous studies have suggested for various reasons that the most appropriate value to take is \cite{Godazgar:2018dvh, Godazgar2019}
\begin{equation}
\lambda = -1.
\end{equation}
With this choice, we can write a first law of the form
\begin{equation}
    \delta \mathcal{M} = \mathcal{T}\delta\mathcal{S} - \frac{3}{2 r_+^2} \delta (n^2 r_+),
\end{equation}
where
\begin{gather}
\mathcal{M}=m-i n, \qquad \mathcal{T} = \frac{1}{4\pi r_+} + \frac{i}{2 \pi} \, \frac{n}{r_+^2}, \qquad \mathcal{S} = \pi r_+^2 - 4\pi i \, n r_+.
\end{gather}
As before, the various components of this equation would need to have a physical interpretation, which remains unclear to us.  However, it is worth speculating whether such a meaningful complex relation is possible.

\paragraph{Acknowledgements} M. G. is supported by a Royal Society University Research Fellowship.

\appendix

\section{First law for the full AdS case}\label{general}

The first law for the AdS case will be a bit more subtle since the action used here contains topological terms that will not affect the global first law but will make the identification of each thermodynamic quantity a bit more difficult. The free energy, which is given by the on-shell action divided by $\beta=1/T$, is on the electric side
\begin{equation}\label{freeenergy}
    F(\beta) = m - TS - \psi N-\frac{\Lambda r_+^3}{8} -3\frac{\left(1-\Lambda n^2\right)^2}{8\Lambda r_+},
\end{equation}
with
\begin{gather}
    T = \frac{1}{4\pi r_+} \left(1-\Lambda(r_+^2+n^2)\right), \quad S_H = \pi \left( r_+^2+n^2\right),\nonumber \\
    \psi = \frac{1}{4\pi n},\quad N = -2\pi \frac{n^3}{r_+}-\frac{4\Lambda}{3}r_+n^3\left( 1-\frac{n^2}{r_+^2}\right),
\end{gather}
where the last term of \eqref{freeenergy} has no nice flat limits, which is a consequence of the form of the action. This should not affect the global first law, but will make the identification of the thermodynamic quantities more subtle for each integral. The idea, however, if one keeps $\Lambda$ constant, should be exactly the same as in the flat case and the topics discussed in this special case thus remain the same.  

In this section, we will simply compute the integrals $\dd Q_\xi\left[\mathcal{H}\right] + 2\dd Q_\xi\left[\mathcal{S}_+\right]$ for the full $n,m,\Lambda\neq\nolinebreak 0$. We will nonetheless consider $\Lambda$ to be non-varying since one would need to deal with new singular terms. The study of the black hole first law with a varying thermodynamic pressure \cite{Cvetic:2010jb} is an interesting topic in its own right that we shall not discuss further here. 

The method used here is exactly the same as in the flat case but the calculations are longer. Let us start with the electric part. 

The horizon integral gives
\begin{align}
    \begin{split}
    \dd Q_\xi\left[\mathcal{H}\right] &= \int_{\mathcal{H}} \delta Q - \iota_\xi\theta \\
    &= \delta m\frac{2 \Lambda n^2 r_+^2+3 n^2 \left(\Lambda n^2-1\right)- \Lambda r_+^4}{ \Lambda \left(n^2+r_+^2\right)^2}\\
    &-n\delta n\frac{11 \Lambda^2 n^6+n^2 \left(-15 \Lambda^2 r_+^4+10 \Lambda r_+^2+6\right)-\Lambda n^4 \left(7 \Lambda r_+^2+17\right)+3 \Lambda  r_+^4 \left(\Lambda r_+^2+1\right)}{2 \Lambda r_+ \left(n^2+r_+^2\right)^2},
    \end{split}
\end{align}
while the Misner string contribution gives 
\begin{align}
    \begin{split}
    2Q_\xi\left[\mathcal{S}_\pm\right]&=\int_{\mathcal{S}_\pm} \delta Q - \iota_\xi\theta\\
    &= n\delta m \frac{ n  \left(3-4 \Lambda r_+^2\right)-4 \Lambda n^3}{ \Lambda \left(n^2+r_+^2\right)^2} \\
    &\hspace{-2mm} +n \delta n\frac{11 \Lambda^2 n^6+n^2 \left(-15 \Lambda^2 r_+^4+10 \Lambda r_+^2+6\right)-\Lambda n^4 \left(7 \Lambda r_+^2+17\right)+3 \Lambda  r_+^4 \left(\Lambda r_+^2+1\right)}{2 \Lambda r_+ \left(n^2+r_+^2\right)^2}
    \end{split},
\end{align}
where we integrated on $\left(r_+,\infty\right)\times (0,2\pi)$ with $\theta=0$ and $t$ kept fixed. It is from these expressions easy to see that the two $\delta n$ terms exactly cancel. Combining the two expressions gives the expected result: 
\begin{equation}
    \dd Q_\xi\left[\mathcal{H}\right]+2Q_\xi\left[\mathcal{S}_\pm\right]=-\delta m 
\end{equation}
The dual charge contribution is computed similarly and gives for the horizon
\begin{align}
    \begin{split}
    \dd \tilde{Q}_\xi\left[\mathcal{H}\right] &= n\delta m\frac{ \left(-3 \Lambda n^4 + n^2  \left(2 \Lambda r_+^2+3\right)+ r_+^2 \left(5 \Lambda r_+^2-3\right)\right)}{2 \Lambda r_+ \left(n^2+r_+^2\right)^2} \\
    &+\frac{\delta n \left(3 \Lambda^2 n^8+n^4 \left(-48 \Lambda^2 r_+^4+58 \Lambda r_+^2+3\right)-2 \Lambda n^6 \left(25 \Lambda r_+^2+3\right)\right.}{4 \Lambda r_+^2 \left(n^2+r_+^2\right)^2}\\
    &\frac{+\left.2 n^2 r_+^2 \left(\Lambda r_+^2 \left(\Lambda r_+^2+11\right)-6\right)-3 r_+^4 \left(\Lambda r_+^2-1\right)^2\right)}{4 \Lambda r_+^2 \left(n^2+r_+^2\right)^2}
    \end{split},
\end{align}
while for the Misner string: 
\begin{align}
    \begin{split}
     2\dd \tilde{Q}_\xi\left[\mathcal{S}_\pm\right] &=n\delta m\frac{\left(3  \Lambda n^4-n^2 \left(2  \Lambda r_+^2+3\right)-r_+^2\left(5 \Lambda r_+^2+3\right)\right)}{2  \Lambda r_+ \left(n^2+\Lambda r_+^2\right)^2}\\
     &+\frac{\delta n \left(3 \Lambda^2 r_+^8+2 \Lambda r_+^6 \left(7 \Lambda n^2-5\right)+r_+^4 \left(10 \Lambda n^2 \left(8 \Lambda n^2-3\right)+3\right)\right.}{4 \Lambda r_+^2 \left(n^2+r_+^2\right)^2}\\
     &\frac{\left.+2 n^2 r_+^2 \left(\Lambda n^2 \left(33 \Lambda n^2-31\right)+6\right)-3 n^4 \left(\Lambda n^2-1\right)^2\right)}{4 \Lambda r_+^2 \left(n^2+r_+^2\right)^2}
     \end{split}.
\end{align}
Observe how the $\delta m$ terms  cancel exactly while the total sum gives
\begin{equation}
     \dd \tilde{Q}_\xi\left[\mathcal{H}\right] + 2\dd \tilde{Q}_\xi\left[\mathcal{S}_\pm\right] = -\delta n \left(1-4\Lambda n^2\right)
\end{equation}
which is nothing but the variation of the dual charge in equation (\ref{charges}). 

\newpage
\bibliographystyle{utphys}
\bibliography{references}

\providecommand{\href}[2]{#2}\begingroup\raggedright\begin{thebibliography}{10}

\bibitem{Abbott:1981ff}
L.~F. Abbott and S.~Deser, ``{Stability of Gravity with a Cosmological
  Constant},'' \href{http://dx.doi.org/10.1016/0550-3213(82)90049-9}{{\em Nucl.
  Phys. B} {\bf 195} (1982)  76--96}.

\bibitem{BH86}
J.~D. Brown and M.~Henneaux, ``{Central charges in the canonical realization of
  asymptotic symmetries: an example from three-dimensional gravity},''
  \href{http://dx.doi.org/cmp/1104114999}{{\em Comm.\ Math.\ Phys.} {\bf 104}
  (1986) no.~2, 207 -- 226}.

\bibitem{Maldacena}
J.~{Maldacena}, \href{http://dx.doi.org/10.1023/A:1026654312961}{``{The Large-N
  Limit of Superconformal Field Theories and Supergravity},''{\em Int.\ J.\
  Theor.\ Phys.} {\bf 38} (Jan., 1999)  1113--1133},
  \href{http://arxiv.org/abs/hep-th/9711200}{{\tt arXiv:hep-th/9711200
  [hep-th]}}.

\bibitem{Godazgar:2018qpq}
H.~Godazgar, M.~Godazgar, and C.~N. Pope, ``{New dual gravitational charges},''
  \href{http://dx.doi.org/10.1103/PhysRevD.99.024013}{{\em Phys. Rev. D} {\bf
  99} (2019) no.~2, 024013}, \href{http://arxiv.org/abs/1812.01641}{{\tt
  arXiv:1812.01641 [hep-th]}}.

\bibitem{Godazgar:2018dvh}
H.~Godazgar, M.~Godazgar, and C.~N. Pope, ``{Tower of subleading dual BMS
  charges},'' \href{http://dx.doi.org/10.1007/JHEP03(2019)057}{{\em JHEP} {\bf
  03} (2019)  057}, \href{http://arxiv.org/abs/1812.06935}{{\tt
  arXiv:1812.06935 [hep-th]}}.

\bibitem{Godazgar:2020gqd}
H.~Godazgar, M.~Godazgar, and M.~J. Perry, ``{Asymptotic gravitational
  charges},'' \href{http://dx.doi.org/10.1103/PhysRevLett.125.101301}{{\em
  Phys. Rev. Lett.} {\bf 125} (2020) no.~10, 101301},
  \href{http://arxiv.org/abs/2007.01257}{{\tt arXiv:2007.01257 [hep-th]}}.

\bibitem{Godazgar:2020kqd}
H.~Godazgar, M.~Godazgar, and M.~J. Perry, ``{Hamiltonian derivation of dual
  gravitational charges},''
  \href{http://dx.doi.org/10.1007/JHEP09(2020)084}{{\em JHEP} {\bf 09} (2020)
  084}, \href{http://arxiv.org/abs/2007.07144}{{\tt arXiv:2007.07144
  [hep-th]}}.

\bibitem{Holst:1995pc}
S.~Holst, ``{Barbero's Hamiltonian derived from a generalized Hilbert-Palatini
  action},'' \href{http://dx.doi.org/10.1103/PhysRevD.53.5966}{{\em Phys. Rev.
  D} {\bf 53} (1996)  5966--5969},
  \href{http://arxiv.org/abs/gr-qc/9511026}{{\tt arXiv:gr-qc/9511026}}.

\bibitem{phaseconserved1}
J.~Lee and R.~M. Wald, ``Local symmetries and constraints,'' {\em J.\ Math.\
  Phys.} {\bf 31} (1990) no.~3, 725--743.

\bibitem{phaseconserved2}
R.~M. {Wald} and A.~{Zoupas},
  \href{http://dx.doi.org/10.1103/PhysRevD.61.084027}{``{General definition of
  ``conserved quantities'' in general relativity and other theories of
  gravity},''{\em Phys.\ Rev.\ D} {\bf 61} (Apr., 2000)  084027},
  \href{http://arxiv.org/abs/gr-qc/9911095}{{\tt arXiv:gr-qc/9911095 [gr-qc]}}.

\bibitem{Aros:1999id}
R.~Aros, M.~Contreras, R.~Olea, R.~Troncoso, and J.~Zanelli, ``{Conserved
  charges for gravity with locally AdS asymptotics},''
  \href{http://dx.doi.org/10.1103/PhysRevLett.84.1647}{{\em Phys. Rev. Lett.}
  {\bf 84} (2000)  1647--1650}, \href{http://arxiv.org/abs/gr-qc/9909015}{{\tt
  arXiv:gr-qc/9909015}}.

\bibitem{Hollands:2005wt}
S.~Hollands, A.~Ishibashi, and D.~Marolf, ``{Comparison between various notions
  of conserved charges in asymptotically AdS-spacetimes},''
  \href{http://dx.doi.org/10.1088/0264-9381/22/14/004}{{\em Class. Quant.
  Grav.} {\bf 22} (2005)  2881--2920},
  \href{http://arxiv.org/abs/hep-th/0503045}{{\tt arXiv:hep-th/0503045}}.

\bibitem{Marolf2014}
D.~Marolf, W.~Kelly, and S.~Fischetti, {\em Conserved Charges in Asymptotically
  (Locally) AdS Spacetimes},
  \href{http://dx.doi.org/10.1007/978-3-642-41992-8_19}{pp.~381--407}.
\newblock Springer, Berlin, Heidelberg, 2014.

\bibitem{Papadimitriou:2005ii}
I.~Papadimitriou and K.~Skenderis, ``{Thermodynamics of asymptotically locally
  AdS spacetimes},''
  \href{http://dx.doi.org/10.1088/1126-6708/2005/08/004}{{\em JHEP} {\bf 08}
  (2005)  004}, \href{http://arxiv.org/abs/hep-th/0505190}{{\tt
  arXiv:hep-th/0505190}}.

\bibitem{Odak:2021axr}
G.~Odak and S.~Speziale, ``{Brown-York charges with mixed boundary
  conditions},'' \href{http://dx.doi.org/10.1007/JHEP11(2021)224}{{\em JHEP}
  {\bf 11} (2021)  224}, \href{http://arxiv.org/abs/2109.02883}{{\tt
  arXiv:2109.02883 [hep-th]}}.

\bibitem{Ashtekar_1984}
A.~Ashtekar and A.~Magnon,
  \href{http://dx.doi.org/10.1088/0264-9381/1/4/002}{``Asymptotically anti-de
  sitter space-times,''{\em Classical and Quantum Gravity} {\bf 1} (jul, 1984)
  L39--L44}.

\bibitem{Ashtekar2000}
A.~{Ashtekar} and S.~{Das},
  \href{http://dx.doi.org/10.1088/0264-9381/17/2/101}{``{Asymptotically anti-de
  Sitter spacetimes: conserved quantities},''{\em Class. Quantum Gravity} {\bf
  17} (Jan., 2000)  L17--L30}, \href{http://arxiv.org/abs/hep-th/9911230}{{\tt
  arXiv:hep-th/9911230 [hep-th]}}.

\bibitem{Barnich:2001jy}
G.~Barnich and F.~Brandt, ``{Covariant theory of asymptotic symmetries,
  conservation laws and central charges},''
  \href{http://dx.doi.org/10.1016/S0550-3213(02)00251-1}{{\em Nucl. Phys. B}
  {\bf 633} (2002)  3--82}, \href{http://arxiv.org/abs/hep-th/0111246}{{\tt
  arXiv:hep-th/0111246}}.

\bibitem{Barnich:2003qn}
G.~Barnich, F.~Brandt, and K.~Claes, ``{Asymptotically anti-de Sitter
  space-times: Symmetries and conservation laws revisited},''
  \href{http://dx.doi.org/10.1016/S0920-5632(03)02410-1}{{\em Nucl. Phys. B
  Proc. Suppl.} {\bf 127} (2004)  114--117},
  \href{http://arxiv.org/abs/gr-qc/0306112}{{\tt arXiv:gr-qc/0306112}}.

\bibitem{Barnich:2004uw}
G.~Barnich and G.~Compere, ``{Generalized Smarr relation for Kerr AdS black
  holes from improved surface integrals},''
  \href{http://dx.doi.org/10.1103/PhysRevD.73.029904}{{\em Phys. Rev. D} {\bf
  71} (2005)  044016}, \href{http://arxiv.org/abs/gr-qc/0412029}{{\tt
  arXiv:gr-qc/0412029}}. [Erratum: Phys.Rev.D 73, 029904 (2006)].

\bibitem{Frodden:2017qwh}
E.~Frodden and D.~Hidalgo, ``{Surface Charges for Gravity and Electromagnetism
  in the First Order Formalism},''
  \href{http://dx.doi.org/10.1088/1361-6382/aa9ba5}{{\em Class. Quant. Grav.}
  {\bf 35} (2018) no.~3, 035002}, \href{http://arxiv.org/abs/1703.10120}{{\tt
  arXiv:1703.10120 [gr-qc]}}.

\bibitem{Durka:2011yv}
R.~Durka and J.~Kowalski-Glikman, ``{Gravity as a constrained BF theory:
  Noether charges and Immirzi parameter},''
  \href{http://dx.doi.org/10.1103/PhysRevD.83.124011}{{\em Phys. Rev. D} {\bf
  83} (2011)  124011}, \href{http://arxiv.org/abs/1103.2971}{{\tt
  arXiv:1103.2971 [gr-qc]}}.

\bibitem{Durka:2011zf}
R.~Durka, ``{Immirzi parameter and Noether charges in first order gravity},''
  \href{http://dx.doi.org/10.1088/1742-6596/343/1/012032}{{\em J. Phys. Conf.
  Ser.} {\bf 343} (2012)  012032}, \href{http://arxiv.org/abs/1111.0961}{{\tt
  arXiv:1111.0961 [gr-qc]}}.

\bibitem{Johnson:2014xza}
C.~V. Johnson, ``{Thermodynamic Volumes for AdS-Taub-NUT and AdS-Taub-Bolt},''
  \href{http://dx.doi.org/10.1088/0264-9381/31/23/235003}{{\em Class. Quant.
  Grav.} {\bf 31} (2014) no.~23, 235003},
  \href{http://arxiv.org/abs/1405.5941}{{\tt arXiv:1405.5941 [hep-th]}}.

\bibitem{Hennigar2019}
R.~A. {Hennigar}, D.~{Kubiznak}, and R.~B. {Mann},
  \href{http://dx.doi.org/10.1103/PhysRevD.100.064055}{``{Thermodynamics of
  Lorentzian Taub-NUT spacetimes},''{\em Phys.\ Rev.\ D} {\bf 100} (Sept.,
  2019)  064055}, \href{http://arxiv.org/abs/1903.08668}{{\tt arXiv:1903.08668
  [hep-th]}}.

\bibitem{Peng:2020cfy}
J.-J. Peng, C.-L. Zou, and H.-F. Liu, ``{A Komar-like integral for mass and
  angular momentum of asymptotically AdS black holes in Einstein gravity},''
  \href{http://dx.doi.org/10.1088/1402-4896/ac1cd1}{{\em Phys. Scripta} {\bf
  96} (2021) no.~12, 125207}, \href{http://arxiv.org/abs/2008.06733}{{\tt
  arXiv:2008.06733 [gr-qc]}}.

\bibitem{Oliveri:2020xls}
R.~Oliveri and S.~Speziale, ``{A note on dual gravitational charges},''
  \href{http://dx.doi.org/10.1007/JHEP12(2020)079}{{\em JHEP} {\bf 12} (2020)
  079}, \href{http://arxiv.org/abs/2010.01111}{{\tt arXiv:2010.01111
  [hep-th]}}.

\bibitem{CompereRuzziconi}
G.~{Comp{\`e}re}, A.~{Fiorucci}, and R.~{Ruzziconi},
  \href{http://dx.doi.org/10.1088/1361-6382/ab3d4b}{``{The
  {\ensuremath{\Lambda}}-BMS$_{4}$ group of dS$_{4}$ and new boundary
  conditions for AdS$_{4}$},''{\em Class. Quantum Gravity} {\bf 36} (Oct.,
  2019)  195017}, \href{http://arxiv.org/abs/1905.00971}{{\tt arXiv:1905.00971
  [gr-qc]}}.

\bibitem{Fiorucci:2020xto}
A.~Fiorucci and R.~Ruzziconi, ``{Charge algebra in Al(A)dS$_{n}$ spacetimes},''
  \href{http://dx.doi.org/10.1007/JHEP05(2021)210}{{\em JHEP} {\bf 05} (2021)
  210}, \href{http://arxiv.org/abs/2011.02002}{{\tt arXiv:2011.02002
  [hep-th]}}.

\bibitem{MahdiTaubNUT}
H.~{Godazgar}, M.~{Godazgar}, and C.~N. {Pope},
  \href{http://dx.doi.org/10.1016/j.physletb.2019.134938}{``{Taub-NUT from the
  Dirac monopole},''{\em Phys.\ Lett.\ B} {\bf 798} (Nov., 2019)  134938},
  \href{http://arxiv.org/abs/1908.05962}{{\tt arXiv:1908.05962 [hep-th]}}.

\bibitem{Godazgar:2018vmm}
H.~Godazgar, M.~Godazgar, and C.~N. Pope, ``{Subleading BMS charges and fake
  news near null infinity},''
  \href{http://dx.doi.org/10.1007/JHEP01(2019)143}{{\em JHEP} {\bf 01} (2019)
  143}, \href{http://arxiv.org/abs/1809.09076}{{\tt arXiv:1809.09076
  [hep-th]}}.

\bibitem{Godazgar2019}
H.~{Godazgar}, M.~{Godazgar}, and C.~N. {Pope},
  \href{http://dx.doi.org/10.1007/JHEP10(2019)123}{``{Dual gravitational
  charges and soft theorems},''{\em JHEP} {\bf 2019} (Oct., 2019)  123},
  \href{http://arxiv.org/abs/1908.01164}{{\tt arXiv:1908.01164 [hep-th]}}.

\bibitem{Hawking:1998ct}
S.~W. Hawking, C.~J. Hunter, and D.~N. Page, ``{Nut charge, anti-de Sitter
  space and entropy},''
  \href{http://dx.doi.org/10.1103/PhysRevD.59.044033}{{\em Phys. Rev. D} {\bf
  59} (1999)  044033}, \href{http://arxiv.org/abs/hep-th/9809035}{{\tt
  arXiv:hep-th/9809035}}.

\bibitem{Miskovic:2009bm}
O.~Miskovic and R.~Olea, ``{Topological regularization and self-duality in
  four-dimensional anti-de Sitter gravity},''
  \href{http://dx.doi.org/10.1103/PhysRevD.79.124020}{{\em Phys. Rev. D} {\bf
  79} (2009)  124020}, \href{http://arxiv.org/abs/0902.2082}{{\tt
  arXiv:0902.2082 [hep-th]}}.

\bibitem{Araneda:2016iiy}
R.~Araneda, R.~Aros, O.~Miskovic, and R.~Olea, ``{Magnetic Mass in 4D AdS
  Gravity},'' \href{http://dx.doi.org/10.1103/PhysRevD.93.084022}{{\em Phys.
  Rev. D} {\bf 93} (2016) no.~8, 084022},
  \href{http://arxiv.org/abs/1602.07975}{{\tt arXiv:1602.07975 [hep-th]}}.

\bibitem{Aros:2017wun}
R.~Aros, ``{Entropy of a Taub-bolt-AdS spacetime from an improved action
  principle},'' \href{http://dx.doi.org/10.1103/PhysRevD.96.024022}{{\em Phys.
  Rev. D} {\bf 96} (2017) no.~2, 024022},
  \href{http://arxiv.org/abs/1705.01177}{{\tt arXiv:1705.01177 [gr-qc]}}.

\bibitem{Giribet:2018hck}
G.~Giribet, O.~Miskovic, R.~Olea, and D.~Rivera-Betancour, ``{Energy in
  Higher-Derivative Gravity via Topological Regularization},''
  \href{http://dx.doi.org/10.1103/PhysRevD.98.044046}{{\em Phys. Rev. D} {\bf
  98} (2018) no.~4, 044046}, \href{http://arxiv.org/abs/1806.11075}{{\tt
  arXiv:1806.11075 [hep-th]}}.

\bibitem{Giribet:2020aks}
G.~Giribet, O.~Miskovic, R.~Olea, and D.~Rivera-Betancour, ``{Topological
  invariants and the definition of energy in quadratic gravity theory},''
  \href{http://dx.doi.org/10.1103/PhysRevD.101.064046}{{\em Phys. Rev. D} {\bf
  101} (2020) no.~6, 064046}, \href{http://arxiv.org/abs/2001.09459}{{\tt
  arXiv:2001.09459 [hep-th]}}.

\bibitem{Arratia:2020hoy}
E.~Arratia, C.~Corral, J.~Figueroa, and L.~Sanhueza, ``{Hairy Taub-NUT/bolt-AdS
  solutions in Horndeski theory},''
  \href{http://dx.doi.org/10.1103/PhysRevD.103.064068}{{\em Phys. Rev. D} {\bf
  103} (2021) no.~6, 064068}, \href{http://arxiv.org/abs/2010.02460}{{\tt
  arXiv:2010.02460 [hep-th]}}.

\bibitem{Ciambelli:2020qny}
L.~Ciambelli, C.~Corral, J.~Figueroa, G.~Giribet, and R.~Olea, ``{Topological
  Terms and the Misner String Entropy},''
  \href{http://dx.doi.org/10.1103/PhysRevD.103.024052}{{\em Phys. Rev. D} {\bf
  103} (2021) no.~2, 024052}, \href{http://arxiv.org/abs/2011.11044}{{\tt
  arXiv:2011.11044 [hep-th]}}.

\bibitem{Flores-Alfonso:2020nnd}
D.~Flores-Alfonso, R.~Linares, and M.~Maceda, ``{Nonlinear extensions of
  gravitating dyons: from NUT wormholes to Taub-Bolt instantons},''
  \href{http://dx.doi.org/10.1007/JHEP09(2021)104}{{\em JHEP} {\bf 09} (2021)
  104}, \href{http://arxiv.org/abs/2012.03416}{{\tt arXiv:2012.03416 [gr-qc]}}.

\bibitem{Mann:2020wad}
R.~B. Mann, L.~A. Pando~Zayas, and M.~Park, ``{Complement to thermodynamics of
  dyonic Taub-NUT-AdS spacetime},''
  \href{http://dx.doi.org/10.1007/JHEP03(2021)039}{{\em JHEP} {\bf 03} (2021)
  039}, \href{http://arxiv.org/abs/2012.13506}{{\tt arXiv:2012.13506
  [hep-th]}}.

\bibitem{Andrianopoli:2021qli}
L.~Andrianopoli, G.~Giribet, D.~L. D\'\i{}az, and O.~Miskovic, ``{Black holes
  with topological charges in Chern-Simons AdS$_{5}$ supergravity},''
  \href{http://dx.doi.org/10.1007/JHEP11(2021)123}{{\em JHEP} {\bf 11} (2021)
  123}, \href{http://arxiv.org/abs/2106.01876}{{\tt arXiv:2106.01876
  [hep-th]}}.

\bibitem{Rodriguez:2021hks}
N.~H. Rodr\'\i{}guez and M.~J. Rodriguez, ``{First Law for Kerr Taub-NUT AdS
  Black Holes},'' \href{http://arxiv.org/abs/2112.00780}{{\tt arXiv:2112.00780
  [hep-th]}}.

\bibitem{Martelli:2012sz}
D.~Martelli, A.~Passias, and J.~Sparks, ``{The supersymmetric NUTs and bolts of
  holography},'' \href{http://dx.doi.org/10.1016/j.nuclphysb.2013.04.026}{{\em
  Nucl. Phys. B} {\bf 876} (2013)  810--870},
  \href{http://arxiv.org/abs/1212.4618}{{\tt arXiv:1212.4618 [hep-th]}}.

\bibitem{Durka:2019ajz}
R.~Durka, ``{The first law of black hole thermodynamics for Taub-NUT
  spacetime},'' \href{http://arxiv.org/abs/1908.04238}{{\tt arXiv:1908.04238
  [gr-qc]}}.

\bibitem{Frodden:2021ces}
E.~Frodden and D.~Hidalgo, ``{The First Law for the Lorentzian Rotating
  Taub-NUT},'' \href{http://arxiv.org/abs/2109.07715}{{\tt arXiv:2109.07715
  [hep-th]}}.

\bibitem{Ramaswamy}
S.~Ramaswamy and A.~Sen, ``{Dual-mass in general relativity},'' {\em
  J.Math.Phys.} {\bf 22} (1981)  2612.

\bibitem{Freidel:2021qpz}
L.~Freidel and D.~Pranzetti, ``{Gravity from symmetry: duality and impulsive
  waves},'' \href{http://dx.doi.org/10.1007/JHEP04(2022)125}{{\em JHEP} {\bf
  04} (2022)  125}, \href{http://arxiv.org/abs/2109.06342}{{\tt
  arXiv:2109.06342 [hep-th]}}.

\bibitem{Freidel:2021ytz}
L.~Freidel, D.~Pranzetti, and A.-M. Raclariu, ``{Higher spin dynamics in
  gravity and $w_{1 + \infty}$ celestial symmetries},''
  \href{http://arxiv.org/abs/2112.15573}{{\tt arXiv:2112.15573 [hep-th]}}.

\bibitem{Godazgar:2022foc}
M.~Godazgar and G.~Long, ``{Higher derivative asymptotic charges and internal
  Lorentz symmetries},'' \href{http://arxiv.org/abs/2201.07014}{{\tt
  arXiv:2201.07014 [hep-th]}}.

\bibitem{AST_1985__S131__95_0}
C.~Fefferman and C.~R. Graham, ``Conformal invariants,'' in {\em \'Elie Cartan
  et les math\'ematiques d'aujourd'hui - Lyon, 25-29 juin 1984}, no.~S131 in
  Ast\'erisque.
\newblock Soci\'et\'e math\'ematique de France, 1985.

\bibitem{Compere:2018aar}
G.~Comp\`ere and A.~Fiorucci, ``{Advanced Lectures on General Relativity},''
  \href{http://arxiv.org/abs/1801.07064}{{\tt arXiv:1801.07064 [hep-th]}}.

\bibitem{Adami:2021nnf}
H.~Adami, D.~Grumiller, M.~M. Sheikh-Jabbari, V.~Taghiloo, H.~Yavartanoo, and
  C.~Zwikel, ``{Null boundary phase space: slicings, news \& memory},''
  \href{http://dx.doi.org/10.1007/JHEP11(2021)155}{{\em JHEP} {\bf 11} (2021)
  155}, \href{http://arxiv.org/abs/2110.04218}{{\tt arXiv:2110.04218
  [hep-th]}}.

\bibitem{Griffiths:2009dfa}
J.~B. Griffiths and J.~Podolsky,
  \href{http://dx.doi.org/10.1017/CBO9780511635397}{{\em {Exact Space-Times in
  Einstein's General Relativity}}}.
\newblock Cambridge Monographs on Mathematical Physics. CUP, Cambridge, 2009.

\bibitem{Misner63}
C.~W. Misner, ``The flatter regions of newman, unti, and tamburino's
  generalized schwarzschild space,''
  \href{http://dx.doi.org/10.1063/1.1704019}{{\em J.\ Math.\ Phys} {\bf 4}
  (1963) no.~7, 924--937}.

\bibitem{Manko:2005nm}
V.~S. Manko and E.~Ruiz, ``{Physical interpretation of NUT solution},''
  \href{http://dx.doi.org/10.1088/0264-9381/22/17/014}{{\em Class. Quant.
  Grav.} {\bf 22} (2005)  3555--3560},
  \href{http://arxiv.org/abs/gr-qc/0505001}{{\tt arXiv:gr-qc/0505001}}.

\bibitem{Wald:1993nt}
R.~M. Wald, ``{Black hole entropy is the Noether charge},''
  \href{http://dx.doi.org/10.1103/PhysRevD.48.R3427}{{\em Phys. Rev. D} {\bf
  48} (1993) no.~8, R3427--R3431},
  \href{http://arxiv.org/abs/gr-qc/9307038}{{\tt arXiv:gr-qc/9307038}}.

\bibitem{Bordo:2019tyh}
A.~B. Bordo, F.~Gray, R.~A. Hennigar, and D.~Kubiz\v{n}\'ak, ``{Misner
  Gravitational Charges and Variable String Strengths},''
  \href{http://dx.doi.org/10.1088/1361-6382/ab3d4d}{{\em Class. Quant. Grav.}
  {\bf 36} (2019) no.~19, 194001}, \href{http://arxiv.org/abs/1905.03785}{{\tt
  arXiv:1905.03785 [hep-th]}}.

\bibitem{BallonBordo:2019vrn}
A.~Ballon~Bordo, F.~Gray, R.~A. Hennigar, and D.~Kubiz\v{n}\'ak, ``{The First
  Law for Rotating NUTs},''
  \href{http://dx.doi.org/10.1016/j.physletb.2019.134972}{{\em Phys. Lett. B}
  {\bf 798} (2019)  134972}, \href{http://arxiv.org/abs/1905.06350}{{\tt
  arXiv:1905.06350 [hep-th]}}.

\bibitem{Cvetic:2010jb}
M.~Cvetic, G.~W. Gibbons, D.~Kubiznak, and C.~N. Pope, ``{Black Hole Enthalpy
  and an Entropy Inequality for the Thermodynamic Volume},''
  \href{http://dx.doi.org/10.1103/PhysRevD.84.024037}{{\em Phys. Rev. D} {\bf
  84} (2011)  024037}, \href{http://arxiv.org/abs/1012.2888}{{\tt
  arXiv:1012.2888 [hep-th]}}.

\end{thebibliography}\endgroup

\end{document}